\documentstyle[12pt,epsfig,psfrag]{article}
\makeatletter
\def\@cite#1#2{{[{#1}]\if@tempswa\typeout
{IJCGA warning: optional citation argument
ignored: `#2'} \fi}}

\newcount\@tempcntc
\def\@citex[#1]#2{\if@filesw\immediate\write\@auxout{\string\citation{#2}}\fi
  \@tempcnta\z@\@tempcntb\m@ne\def\@citea{}\@cite{\@for\@citeb:=#2\do
    {\@ifundefined
       {b@\@citeb}{\@citeo\@tempcntb\m@ne\@citea\def\@citea{,}{\bf ?}\@warning
       {Citation `\@citeb' on page \thepage \space undefined}}%
    {\setbox\z@\hbox{\global\@tempcntc0\csname b@\@citeb\endcsname\relax}%
     \ifnum\@tempcntc=\z@ \@citeo\@tempcntb\m@ne
       \@citea\def\@citea{,}\hbox{\csname b@\@citeb\endcsname}%
     \else
      \advance\@tempcntb\@ne
      \ifnum\@tempcntb=\@tempcntc
      \else\advance\@tempcntb\m@ne\@citeo
      \@tempcnta\@tempcntc\@tempcntb\@tempcntc\fi\fi}}\@citeo}{#1}}
\def\@citeo{\ifnum\@tempcnta>\@tempcntb\else\@citea\def\@citea{,}%
  \ifnum\@tempcnta=\@tempcntb\the\@tempcnta\else
   {\advance\@tempcnta\@ne\ifnum\@tempcnta=\@tempcntb \else \def\@citea{--}\fi
    \advance\@tempcnta\m@ne\the\@tempcnta\@citea\the\@tempcntb}\fi\fi}
\makeatother
\newenvironment{Eqnarray}%
     {\arraycolsep 0.14em\begin{eqnarray}}{\end{eqnarray}}

\def\be{\begin{equation}}
\def\ee{\end{equation}}
\def\bear{\be\begin{array}}
\def\eear{\end{array}\ee}
\def\bea{\begin{Eqnarray}}
\def\eea{\end{Eqnarray}}

\def\lsim{\mathrel{\raise.3ex\hbox{$<$\kern-.75em\lower1ex\hbox{$\sim$}}}}
\def\gsim{\mathrel{\raise.3ex\hbox{$>$\kern-.75em\lower1ex\hbox{$\sim$}}}}
\def\ifmath#1{\relax\ifmmode #1\else $#1$\fi}
\def\ls#1{\ifmath{_{\lower1.5pt\hbox{$\scriptstyle #1$}}}}

\def\beq{\begin{equation}}
\def\eeq{\end{equation}}
\def\beqa{\begin{Eqnarray}}
\def\eeqa{\end{Eqnarray}}

\def\snu{\tilde{\nu}}
\def\gappeq{\mathrel{\rlap {\raise.5ex\hbox{$>$}}
{\lower.5ex\hbox{$\sim$}}}}
\def\lappeq{\mathrel{\rlap{\raise.5ex\hbox{$<$}}
{\lower.5ex\hbox{$\sim$}}}}

\begin{document}
\def\IJMPA #1 #2 #3 {{\sl Int.~J.~Mod.~Phys.}~{\bf A#1}\ (19#2) #3$\,$}
\def\MPLA #1 #2 #3 {{\sl Mod.~Phys.~Lett.}~{\bf A#1}\ (19#2) #3$\,$}
\def\NPB #1 #2 #3 {{\sl Nucl.~Phys.}~{\bf B#1}\ (19#2) #3$\,$}
\def\PLB #1 #2 #3 {{\sl Phys.~Lett.}~{\bf B#1}\ (19#2) #3$\,$}
\def\PR #1 #2 #3 {{\sl Phys.~Rep.}~{\bf#1}\ (19#2) #3$\,$}
\def\JHEP #1 #2 #3 {{\sl JHEP}~{\bf #1}~(19#2)~#3$\,$}
\def\PRD #1 #2 #3 {{\sl Phys.~Rev.}~{\bf D#1}\ (19#2) #3$\,$}
\def\PTP #1 #2 #3 {{\sl Prog.~Theor.~Phys.}~{\bf #1}\ (19#2) #3$\,$}
\def\PRL #1 #2 #3 {{\sl Phys.~Rev.~Lett.}~{\bf#1}\ (19#2) #3$\,$}
\def\RMP #1 #2 #3 {{\sl Rev.~Mod.~Phys.}~{\bf#1}\ (19#2) #3$\,$}
\def\ZPC #1 #2 #3 {{\sl Z.~Phys.}~{\bf C#1}\ (19#2) #3$\,$}
\def\PPNP#1 #2 #3 {{\sl Prog. Part. Nucl. Phys. }{\bf #1} (#2) #3$\,$}

\catcode`@=11
\newtoks\@stequation
\def\subequations{\refstepcounter{equation}%
\edef\@savedequation{\the\c@equation}%
  \@stequation=\expandafter{\theequation}
  \edef\@savedtheequation{\the\@stequation}
  \edef\oldtheequation{\theequation}%
  \setcounter{equation}{0}%
  \def\theequation{\oldtheequation\alph{equation}}}
\def\endsubequations{\setcounter{equation}{\@savedequation}%
  \@stequation=\expandafter{\@savedtheequation}%
  \edef\theequation{\the\@stequation}\global\@ignoretrue

\noindent}
\catcode`@=12

\vspace*{-1in}
\renewcommand{\thefootnote}{\fnsymbol{footnote}}
\begin{flushright}
OUTP-02-30P \\
IPPP/02/33 \\
DCPT/02/66
\end{flushright}
\vskip 5pt
\begin{center}
{\Large {\bf Leptogenesis and low-energy phases }}
\vskip 25pt
{\bf Sacha Davidson$^*$ and
Alejandro Ibarra$^{\dagger}$}
\vskip 10pt 
 {\it $^*$ Dept of Physics, University of Durham,
Durham, DH1 3LE, UK}\\
{\it $^{\dagger}$ Theoretical Physics, University of Oxford,
 1 Keble Road, Oxford, OX1 3NP, UK}\\
\vskip 20pt
{\bf Abstract}
\end{center}
\begin{quotation}
  {\noindent\small 
In supersymmetric  models,
the CP asymmetry produced in the decay of the
lightest
right-handed neutrino, $\equiv \epsilon$,
can be written as a function of weak scale parameters.
We introduce a way of separating
$\epsilon$ into contributions from  the
various weak-scale phases, and study the contribution
of potentially measurable neutrino phases to leptogenesis.
We find that the Majorana
phase $\phi'$, which could have observable effects
on neutrinoless double beta decay,
is important for  $\epsilon$ unless there
are cancellations among phases.  If  the phase
$\delta$  can be measured at a neutrino factory, 
then it contributes significantly
to $\epsilon$ over much of parameter space.
\vskip 10pt
\noindent
}

\end{quotation}

\vskip 20pt  

\setcounter{footnote}{0}
\renewcommand{\thefootnote}{\arabic{footnote}}

\section{Introduction}

After the discovery of neutrino oscillations\cite{Cleveland:1998nv,SK},
 leptogenesis
\cite{Fukugita:1986hr}
stands as one of the most appealing explanations for the observed
Baryon Asymmetry of the Universe (BAU)
\cite{Buchmuller:2000wq}. One of the crucial ingredients 
 \cite{Sakharov:1967dj} for this mechanism is CP violation 
in the leptonic sector. However, 
there is no indication for it so far, 
although it could perhaps be seen at a neutrino
factory or in neutrinoless double beta decay. 
It is therefore interesting to investigate whether
there is any relation between the CP violation
required for leptogenesis and the phases that
could be measured at low energies in the neutrino sector.

The leptogenesis scenario relies on the seesaw model \cite{seesaw} 
for neutrino masses, that is usually analyzed in terms of
high-energy parameters, not accessible to experiments. So, the
resulting predictions are (texture) model-dependent.
The above question has been addressed in such an approach 
\cite{Branco:2001pq,Branco:2002kt,Ellis:2001xt,LR,GUT,King:2002nf}.
Instead, we  parametrize the seesaw in terms of
weak scale variables \cite{Davidson:2001zk}. This  gives us a
model-independent formulation of leptogenesis in terms of low energy
inputs, in which we can study the above question.

The aim of this paper is to quantify, in a model
independent way,  the relation
of the CP violation required for leptogenesis to the
measurable low-energy phases.  We express the
CP asymmetry of leptogenesis as a function of 
real parameters and phases at the weak scale,
and then introduce a definition of ``phase overlap''
between the leptogenesis phase and the individual
low energy phases. This  definition   is not the only
possible one,
but has linearity properties and is calculable. 
It is motivated by the notion  of vector space,
spanned by low energy phases (``basis vectors''),
in which the CP asymmetry of leptogenesis is 
a ``vector''. The relative importance of a low
energy phase for leptogenesis would then be
the ``inner product'' of the leptogenesis ``vector''
with the relevant ``basis vector''. 
 We will not be able to construct such a vector
space, but it is a useful analogy to keep in mind. 

The paper is organized as follows: the next
section introduces CP violation in the leptonic sector. 
Section \ref{3} contains the
basic concepts of the supersymmetric see-saw and the generation
of the BAU  by leptogenesis, from a top-down point of view. 
In Section \ref{4}   we  review the procedure 
to reformulate the see-saw mechanism from a bottom-up
perspective. This will allow us to study leptogenesis
in terms of low energy data, opening the possibility
of relating, in a straight-forward way, the Baryon Asymmetry
of the Universe with the CP violation measurable at neutrino factories. 
In section \ref{5} we develop the general formalism to study quantitatively
the above-mentioned relation. In Section \ref{6} and \ref{7} 
we show the results of
our analysis, first for a particular case, and then 
for a more general case.  In section 
\ref{9} we present a self-contained summary
and conclusions. Finally, we include an appendix
with the  procedure to evaluate numerically
the contributions from the low-energy phases to leptogenesis.

\section{Flavour and CP violation in the leptonic sector}

In the last few years, the Superkamiokande collaboration \cite{SK}
has provided compelling evidence that neutrinos have mass
and oscillate. More recently, the SNO collaboration \cite{SNOCC} has
confirmed the oscillation hypothesis, and the first neutral
current data \cite{Ahmad:2002jz} seem to favour the large angle
MSW (LAMSW) solution to the solar neutrino problem \cite{Bahcall:2002hv}. 
 These results, 
combined with those from a series of other experiments \cite{experiments}, 
have allowed
to measure fairly well the mass splittings and mixing
angles relevant for solar and atmospheric 
neutrino oscillations. In addition
to this, other experiments have provided bounds on neutrino parameters
from electron antineutrino disappearance (CHOOZ)\cite{Apollonio:1999ae}, 
the non-observation
of neutrinoless double beta decay \cite{0nbb}, the shape of the tritium beta
decay spectrum \cite{Lobashev:tp}, and different cosmological
and astrophysical considerations. However, no evidence has been 
found so far for CP violation in the leptonic sector. 

The search for leptonic CP violation  
is theoretically motivated by several facts.
First, the discovery of CP violation in the leptonic sector
could shed some light on the mechanism that generates 
neutrino masses and perhaps hint at some underlying structure.
Secondly, the observation of CP violation in 
the quark sector, (in the neutral kaon system, $\epsilon'/\epsilon$,
and in $B \rightarrow \psi K_s$),  encourages the search
for CP violation in the neutrino sector. If there exists
a symmetry relating quarks and leptons, these experimental results
would point to CP violation also in the leptonic sector. 
Furthermore, particular models 
 would give definite predictions
that could be contrasted in the future.
Lastly, and most importantly for the purposes of this paper, 
CP violation in the leptonic sector could
be related to the observed Baryon Asymmetry of 
the Universe. This is possible in the context of 
the see-saw mechanism.

On the experimental side, the leptonic version of the CKM phase
can be detected
by comparing transition probabilities
for neutrinos and antineutrinos:
\bea
A=\frac{
P(\nu_{\alpha} \rightarrow \nu_{\beta})-
P(\bar \nu_{\alpha} \rightarrow \bar \nu_{\beta})}
{P(\nu_{\alpha} \rightarrow \nu_{\beta})+
P(\bar \nu_{\alpha} \rightarrow \bar \nu_{\beta})} ~~.
\eea
Unfortunately, it is not possible to measure such an  asymmetry
with the natural sources of neutrinos, i.e. the Sun and 
pions decaying in the atmosphere, since the ``beam'' cannot
be switched from $\nu$ to $\bar{\nu}$.
 Hence, a lot of effort
is being bestowed on the design of a neutrino factory 
\cite{nufact,Cervera:2000kp}:
  an intense muon source to produce
a high-intensity neutrino beam. In the muon storage 
ring, muons decay to produce muon neutrinos
and electron antineutrinos. Whereas a muon neutrino would produce
a muon in the detector, the oscillation of an electron antineutrino
to a muon antineutrino would produce an antimuon. This antimuon
(a ``wrong sign'' muon) would be a clear signature for
oscillation, and $P(\bar \nu_{e} \rightarrow \bar \nu_{\mu})$
could be determined. One of the advantages of a neutrino factory
is that the muons in the storage ring can be replaced by antimuons.
This makes possible the measurement of  
$P(\nu_{e} \rightarrow \nu_{\mu})$ and hence the CP asymmetry.
In practice, detecting CP violation in the neutrino sector is not
an easy task \cite{Romanino:1999zq,Cervera:2000kp}, 
since the beam has to go through the Earth, that
is CP asymmetric. In consequence, the matter effects on the
oscillation pattern can obscure the CP violation intrinsic to 
neutrinos.

If neutrinos have Majorana masses, as predicted by the seesaw mechanism,
there are also ``Majorana'' phases, in addition to
the ``Dirac'' phase that could be detected at a neutrino
factory.  Neutrinoless double beta decay could
be sensitive to these phases (see however
\cite{Barger:2002vy}). This lepton number
violating, but CP conserving, process probes
the Majorana neutrino mass matrix element
between $\nu_e$ and $\nu_e$,
which     depends on the masses and mixing angles,
and also on the Majorana phases. Neutrinoless double
beta decay is not observed at the moment. However,
experiments which should see a signal, for  the currently
favored masses and mixing angles (LMA), are being discussed
 \cite{Elliott:2002xe}.

 The see-saw mechanism \cite{seesaw} consists 
on adding three right-handed neutrinos to the 
Standard Model (SM) particle content, singlets with respect to the
SM gauge group, and coupled to the Higgs doublet through a
Yukawa coupling. Then, a Majorana mass term for the right-handed 
neutrinos is not forbidden by the gauge symmetry, and can be
naturally much larger than the scale of electroweak symmetry breaking.
These simple assumptions are enough to produce neutrino masses
naturally small \footnote{Nevertheless, this minimal model has 
a serious hierarchy problem: the right-handed neutrinos
produce a (large) quadratically divergent radiative correction
to the Higgs mass. Therefore, in what follows, we will
restrict ourselves to the supersymmetric version of the see-saw
mechanism.}. Furthermore, if CP is violated in the leptonic
sector, the decay of the right-handed neutrinos in the early Universe
produces a lepton asymmetry  \cite{Fukugita:1986hr,Buchmuller:1999cu}
that will be eventually reprocessed into a baryon asymmetry
by sphalerons \cite{Kuzmin:1985mm}. This  leptogenesis scenario
 will be discussed in more detail in section \ref{3}.

In supersymmetric models, the seesaw mechanism 
can induce  flavour violating processes 
involving charged leptons that could be observed in the
future \cite{Borzumati:1986qx}.
The neutrino Yukawa couplings generate
off-diagonal elements in the 
slepton mass matrix,  via renormalization
group running.  These flavour violating mass terms contribute inside
 loops  to  processes
such as $\mu \rightarrow e \gamma$, 
 $\tau \rightarrow \mu \gamma$ and 
 $\tau \rightarrow e \gamma$.
This has been extensively studied from various
theoretical \cite{topdown,King:2002nf}  and   phenomenological
\cite{bottomup,Lavignac:2001vp,Casas:2001sr} perspectives.
The current experimental bound \cite{Brooks:1999pu}
on $\mu \rightarrow e \gamma$
imposes some restrictions on the parameter space
of the SUSY seesaw. It is anticipated that
the sensitivity to   $\tau \rightarrow \mu \gamma$
and  $\mu \rightarrow e \gamma$ could improve by
as much as three orders of magnitude \cite{prop} in forthcoming years.
This would provide interesting information about
the flavour structure of the SUSY seesaw, irrespective of
whether lepton flavour violation is observed or not.

%
%

In this paper, we will concentrate on the possible
relation of the CP asymmetry in the leptonic sector with 
the Baryon Asymmetry of the Universe, in the framework of
the supersymmetric leptogenesis. We suppose that the
BAU is generated in the out-of-equilibrium decay of
the {\it lightest} right-handed neutrino.
The CP violation that gives rise to the
BAU is not straight-forwardly related to the CP violation
that could be observed at low energy.
This has been carefully and elegantly discussed,
using Jarlskog invariants in \cite{Branco:2001pq}. These authors
have also studied the relation of the leptogenesis
phase to low energy phases for specific high scale models and various
solar solutions \cite{Branco:2002kt}. They have an analytic 
approximation similar to ours, and our results seem to
agree where  they overlap.   Correlations between
leptogenesis and low energy parameters have been studied
in left-right symmetric models where the Yukawa couplings are
small \cite{LR}, and in various Grand Unified Theories \cite{GUT}.
In certain classes of top-down models,
it has been found   \cite{King:2002nf} that the leptogenesis
phase is unrelated to the MNS phases; 
these textures  correspond to the third case we study, in section
\ref{7.2}, so the conclusion  \cite{King:2002nf}
that $\delta$ makes little contribution to leptogenesis
agrees with our result. 
The goal of this paper is to investigate the interplay between the
CP violation at very high energies and at low energies, 
in a model independent way. We will also comment on the
prospects to observe CP violation at  a neutrino factory
or in neutrinoless double beta decay, in view of the
measured BAU, and inversely, what could be inferred about the
BAU if CP violation is observed at low energy.

\section{The see-saw mechanism and leptogenesis: the top-down approach}
\label{3}

The supersymmetric version of the see-saw mechanism has a 
leptonic superpotential that reads
\bea
\label{superp}
W_{lep}= {e_R^c}^T {\bf Y_e} L\cdot H_d 
+ {\nu_R^c}^T {\bf Y_\nu} L\cdot H_u 
- \frac{1}{2}{\nu_R^c}^T{\cal M}\nu_R^c , \eea
where $L_i$ and $e_{Ri}$ ($i=e, \mu, \tau$) are the left-handed 
lepton doublet and the right-handed charged-lepton singlet, 
respectively, and $H_d$ ($H_u$) is the hypercharge $-1/2$ ($+1/2$)
 Higgs doublet.
 ${\bf Y_e}$ and ${\bf Y_{\nu}}$ are the Yukawa couplings that 
give masses to the charged leptons and generate the neutrino Dirac mass, 
and $\cal M$ is a $3 \times 3$ Majorana mass matrix that does 
not break the SM gauge symmetry. We do not  make any 
assumptions about the structure of the matrices in eq.(\ref{superp}), 
but consider the most general case. Then, it can be proved 
that the number of independent physical parameters is 21: 
15 real parameters and 6  phases \cite{Santamaria:1993ah}.

It is natural to assume that the  overall scale of $\cal M$, 
denoted by $M$, is much larger than the electroweak scale or any soft mass. 
Therefore, at low energies the right-handed neutrinos are decoupled and 
the corresponding effective Lagrangian reads
\bea  \delta {\cal L}_{lep}={e_R^c}^T {\bf Y_e} L\cdot H_d 
-\frac{1}{2}({\bf Y_\nu}L\cdot H_u)^T{\cal
M}^{-1}({\bf Y_\nu}L\cdot H_u) + {\rm h.c.}.
\eea
So, after the electroweak symmetry breaking, the left-handed neutrinos
acquire a Majorana mass, given by
\bea
\label{seesaw}
{\cal M}_\nu= {\bf m_D}^T {\cal M}^{-1} {\bf m_D}, \eea
suppressed with respect to the typical fermion masses by the inverse 
power of the large scale $M$.

 We will find convenient
to work in the flavour basis  where the charged-lepton Yukawa
matrix, $\bf{Y_e}$, and the gauge interactions are flavour-diagonal.
In this basis, the neutrino mass matrix, ${\cal M}_\nu$, 
can be diagonalized by the MNS \cite{Maki:1962mu} matrix $U$, 
defined by
\be
\label{Udiag}
U^T {\cal M}_\nu U={\mathrm diag}(m_{\nu_1},m_{\nu_2},m_{\nu_3})\equiv
D_{ {\cal M}_\nu }, \ee
where $U$ is a unitary matrix that relates  flavour to mass eigenstates
\bea  \pmatrix{\nu_e \cr \nu_\mu\cr \nu_\tau\cr}= U \pmatrix{\nu_1\cr
\nu_2\cr \nu_3\cr}\,,
\label{CKM}
\eea
and the $m_{\nu_i}$ can be chosen real and positive. Also, we label
the masses in such a way that $m_{\nu_1}<m_{\nu_2}<m_{\nu_3}$. 
We will assume throughout the paper that the light neutrinos
have a hierarchical spectrum. We do not consider the inverse
hierarchy, which may be more difficult to match with
the neutrinos detected from SN1987A \cite{Dighe:1999bi}; we anticipate
that the contribution of $\delta$ to the leptogenesis phase
could be suppressed by powers of $m_{\nu_1}$ in this case. 
The CP asymmetry required for leptogenesis is suppressed for
degenerate light neutrinos \cite{Davidson:2002qv}, 
so we neglect this possibility.

$U$ can be written as
\bea U=V\cdot {\mathrm diag}(e^{-i\phi/2},e^{-i\phi'/2},1)\ \ ,
\label{UV}
\eea
where $\phi$ and $\phi'$ are CP violating phases (if different from
$0$ or $\pi$) and $V$ has the ordinary form of the CKM matrix
\be 
\label{Vdef} 
V=\pmatrix{c_{13}c_{12} & c_{13}s_{12} & s_{13}e^{-i\delta}\cr
-c_{23}s_{12}-s_{23}s_{13}c_{12}e^{i\delta} & c_{23}c_{12}-s_{23}s_{13}s_{12}e^{i\delta} & s_{23}c_{13}\cr
s_{23}s_{12}-c_{23}s_{13}c_{12}e^{i\delta} & -s_{23}c_{12}-c_{23}s_{13}s_{12}e^{i\delta} &
c_{23}c_{13}\cr}. 
\ee
It is interesting to note that the neutrino mass matrix, 
${\cal M}_\nu$, depends on 9 parameters: 6 real parameters and
3 phases. Comparing with the complete theory, we discover that
some information has been ``lost'' in the decoupling process, to be
precise, 6 real parameters and three phases. We will return to this
important issue later on.

Another remarkable feature of the see-saw mechanism is
that it provides a natural framework to generate the baryon
asymmetry of the Universe, defined by 
$\eta_B = (n_B - n_{\bar B})/s$, where $s$ is the entropy
density. This quantity is strongly constrained by Big Bang
Nucleosynthesis to lie in the range 
$\eta_B \simeq (0.3-0.9) \times 10^{-10}$, to successfully
reproduce the observed abundances of the light nuclei 
D, $^3$He, $^4$He and $^7$Li \cite{Olive:2000ij}. As was shown by Sakharov, 
generating a baryon asymmetry requires baryon number violation, 
C and CP violation, and a deviation from thermal equilibrium.
These three conditions are fulfilled in the out-of-equilibrium
decay of the right-handed neutrinos and sneutrinos in the early 
Universe. For conciseness, and since we
are concerned only with supersymmetric leptogenesis, in what follows
we will use right-handed neutrinos, and the shorthand notation
$\nu_R$, to refer both to right-handed neutrinos and right-handed
sneutrinos.

Let us briefly review the mechanism of generation of the BAU through
leptogenesis \cite{Fukugita:1986hr,Buchmuller:1999cu}.
At the end of inflation, a certain number density of right-handed
neutrinos, $n_{\nu_R}$, is produced, that depends on the cosmological
scenario. These right-handed neutrinos decay, with a decay rate that
reads, at tree level,
\beq
\Gamma_{D_i}= \Gamma(\nu_{R_i} \rightarrow  \ell_i H_u) + \Gamma
({\nu}_{R_i} \rightarrow  \tilde{L}_i \tilde{h}_u)
= \frac{1}{8 \pi} ({\bf Y_{\nu} Y_{\nu}}^{\dagger})_{ii} M_i.
\label{decay-rate}
\eeq
The out of equilibrium decay of a right-handed neutrino
$\nu_{R_i}$ creates a lepton asymmetry given by
\beq
\eta_{L} = \frac{n_\ell - n_{\bar \ell}}{s} = 
 \frac{n_{\nu_R} +n_{\tilde{\nu}_R} }{s} ~ \epsilon_i ~ \kappa.
\label{etaL}
\eeq
The value of $(n_{\nu_R} +n_{\tilde{\nu}_R})/s$ depends on the
particular mechanism to generate the right-handed (s)neutrinos. On the
other hand, the CP-violating parameter
\beq
\epsilon_i = \frac{\Gamma_{D_i}- \bar \Gamma_{D_i}}
{\Gamma_{D_i}+ \bar \Gamma_{D_i}},
\label{epsilon}
\eeq
where $\bar \Gamma_{D_i}$ is the CP conjugated version of
$\Gamma_{D_i}$, is determined by the particle physics model that gives
the masses and couplings of the $\nu_R$.  Finally, $\kappa$ is the
fraction of the produced asymmetry that survives washout by lepton
number violating interactions after $\nu_R$ decay.  To ensure $\kappa
\sim 1$, lepton number violating interactions (decays, inverse decays
and scatterings) must be out of equilibrium when the right-handed
neutrinos decay. In the case of
the lightest right-handed neutrino  $\nu_{R_1}$,
this corresponds approximately to $\Gamma_{D_1}<H|_{T
\simeq M_1}$, where $H$ is the Hubble parameter at the temperature
$T$, and can be expressed in terms of an effective light neutrino mass
\cite{Buchmuller:1999cu,Plumacher:1997kc}, 
$\widetilde m_1$,  as
\beq
\widetilde{m}_1 = 
\frac{ 8 \pi \langle H_u^0\rangle ^2}{M_1^2}
\Gamma_{D_1} = 
({\bf Y_{\nu} Y_{\nu}}^{\dagger})_{11} 
\frac{\langle H_u^0\rangle ^2}{M_1}
\lappeq 5 \times 10^{-3} {\rm eV}.
\label{mtilde}
\eeq
This requirement has been carefully studied
\cite{Buchmuller:1999cu,Plumacher:1997kc,Barbieri:2000ma}; the precise
numerical bound on $\widetilde m_1$ depends on $M_1$, and can be found
in \cite{Plumacher:1997kc}.

The last step is the transformation of the lepton asymmetry into a
baryon asymmetry by non-perturbative B+L violating (sphaleron)
processes
\cite{Kuzmin:1985mm}, giving 
\beq
\eta_{B} =\frac{C}{C-1} \eta_{L},
\label{BAU}
\eeq
where $C$ is a number ${\cal O}(1)$, that in the Minimal
Supersymmetric Standard Model takes the value $C=8/23$.

In this paper, we assume  that a sufficient number
of $\nu_R$ were produced --- thermally, or
 in the decay of the inflaton,
or as a scalar condensate of $\tilde{\nu}_R$s, 
or by some other mechanism.
 We will concentrate on the step of leptogenesis
that is most directly related to neutrino physics, 
namely the generation of a CP asymmetry, $\epsilon$, in the
decay of the right handed neutrinos.
It is convenient to work in a basis of 
right-handed neutrinos where ${\cal M}$ is diagonal
\be {\cal M}={\rm{diag}} (M_1,M_2,M_3),
\ee
with $M_i$ real and $0 \leq M_1 <M_2 <M_3$. 
In this basis, the CP asymmetry 
produced in the decay of $\nu_{R_i}$ reads
\bea
\epsilon_i  \simeq  - \frac{1}{8 \pi} \frac{1}{[{\bf Y_{\nu} Y_{\nu}}
^{\dagger}]_{ii}}  \sum_j {\rm Im}
 \left\{ [{\bf Y_{\nu} Y_{\nu}}^{\dagger}]^2_{ij} \right\}
f \left(  \frac{M_j^2}{M_i^2} \right),
\eea
where \cite{Covi:1996wh}
\beq
f(x) = \sqrt{x}  \left( \frac{2}{x - 1 } + \ln \left[ \frac{1+x}{x} \right]
\right).
\eeq
Here, we will assume that the masses of the right-handed neutrinos are
hierarchical. We also assume that 
the lepton asymmetry is  generated in the decay of the
lightest right-handed neutrino.  
This second assumption is critical; if
the asymmetry was generated by the decay of
$\nu_{R_2}$ or $\nu_{R_3}$, it would depend on
a different combination of  phases.
 This assumption is also dubious if
 the $\nu_R$ are produced  thermally,  because
the $\nu_{R_1}$ mass, $M_1$, is severely constrained
in SUSY models. To get a large
enough baryon asymmetry, $M_1 > 10^8$ GeV is required
\cite{Davidson:2002qv,Hamaguchi:2001gw}, but $M_1$ must be less than or of 
order the reheat temperature $T_{reh}$. To avoid overproducing
gravitons in the early Universe,  $T_{reh}$ is required to be
$ \lsim 10^{9}- 10^{10}$ GeV \cite{Kawasaki:1995af}. 

With these approximations, the CP asymmetry is
\bea
\epsilon_1 &\simeq&  -\frac{3}{8 \pi} \frac{1}{[{\bf Y_{\nu} Y_{\nu}}
^{\dagger}]_{11}} \sum_j {\rm {Im}} \left\{ [{\bf Y_{\nu}
Y_{\nu}^{\dagger}}]^2_{1j} \right\} \left( \frac{M_1}{M_j} \right) \\
&=& - \frac{3}{8 \pi \langle H_u^0\rangle ^2}\frac{M_1}
{[{\bf Y_{\nu} Y_{\nu}}^{\dagger}]_{11}} {\rm {Im}} \left\{ [{\bf
Y_{\nu}} {\cal M_{\nu}}^{\dagger} {\bf Y_{\nu}}^T]_{11} \right\}.
\label{eps1}
\eea

The CP asymmetry depends on quantities that appear in the
superpotential of the complete theory, eq.(\ref{superp}),
and that are not directly measurable with experiments. 
However, these quantities can be related to neutrino and
sneutrino parameters, as we will discuss in the next section.
One of the goals of this paper is to implement in an explicit way
these constraints on the CP asymmetry, eq.(\ref{eps1}).

\section{The see-saw mechanism and leptogenesis: the bottom-up approach}
\label{4}

Our starting point will be the procedure presented in 
\cite{Davidson:2001zk}. 
In the basis defined in section \ref{3}, where the charged
lepton mass matrix and the right-handed Majorana mass matrix
are diagonal, the neutrino Yukawa coupling must be necessarily
non-diagonal. However, it can be diagonalized by two
unitary transformations:
\beq
\label{biunitary}
{\bf Y_\nu} = V_R^{\dagger} D_Y V_L.  \eeq It is clear that the CP
asymmetry depends just on $V_R$ and $D_{\cal M}$.  These quantities are
related to the physics of the right-handed neutrinos and are not directly
testable by experiments, since they are related to very high energy
physics. However, there is a reminiscence of $V_R$ and
$D_{\cal M}$ in the low energy neutrino mass matrix that can be
exploited to obtain information about the high-energy physics from the
neutrino data. Substituting eq.(\ref{biunitary}) in the see-saw
formula, ${\cal M}_\nu= {\bf m_D}^T {\cal M}^{-1} {\bf m_D}$, one
obtains\beq
\label{see-saw2}
D_Y^{-1} V_L^* \frac{{\cal M}_\nu}{\langle H_u^0\rangle ^2}
 V^{\dagger}_L D_Y^{-1} 
 = V_R^*  D_{\cal M}^{-1} V^{\dagger}_R  \equiv {\cal M}^{-1}.
\eeq
From this equation we can solve for $V_R$
and $D_{\cal M}$ in terms of ${\cal M}_{\nu}$, $D_Y$ and $V_L$.
${\cal M}_{\nu}$ is constrained by neutrino experiments, whereas 
$D_Y= {\mathrm diag}(y_1,y_2,y_3)$ and $V_L$ are unknown
parameters at this stage. We choose a parametrization 
of the unitary matrix $V_L$ such that
\be 
\label{VLdef} 
V_L=\pmatrix{c^L_{13}c^L_{12} & 
c^L_{13}s^L_{12} e^{-i \varphi_{12}} & 
s^L_{13}e^{-i\varphi_{13}} \cr
-c^L_{23}s^L_{12} e^{i \varphi_{12}}
-s^L_{23}s^L_{13}c^L_{12}e^{i(\varphi_{13}-\varphi_{23})} &
 c^L_{23}c^L_{12}-s^L_{23}s^L_{13}s^L_{12}e^{-i(\varphi_{12}
-\varphi_{13}+\varphi_{23})} & 
s^L_{23}c^L_{13} e^{-i\varphi_{23}}\cr
s^L_{23}s^L_{12}e^{i(\varphi_{12}+\varphi_{23})}
-c^L_{23}s^L_{13}c^L_{12}e^{i\varphi_{13}} &
 -s^L_{23}c^L_{12}e^{i \varphi_{23}}
-c^L_{23}s^L_{13}s^L_{12}e^{i(\varphi_{13}-\varphi_{12})} &
c^L_{23}c^L_{13}\cr},
\ee
where $c_{ij}^L=\cos \theta_{ij}^L$ and $s_{ij}^L=\sin \theta_{ij}^L$,
being $\theta_{ij}^L$ the angles in the $V_L$ matrix.

In certain scenarios,  the parameters $D_Y$ and
$V_L$ can be constrained experimentally. For example, in a scenario
of minimal SUGRA, with just the MSSM+3$\nu_R$s below the GUT
scale,  $D_Y$ and $V_L$ can in principle  be extracted from the
radiative corrections to the left-handed slepton mass matrix,
since the corresponding RGE depends on the combination 
${\bf Y^{\dagger}_{\nu}} {\bf Y_\nu}= V_L^{\dagger} D_Y^2 V_L$.
To be more precise, at low energies the left-handed slepton mass matrix 
reads, in the leading-log approximation
\bea
\label{softafterRG} 
\left(m^2_{\tilde{\ell}, \snu}\right)_{ij} & \simeq & 
 \left({\rm diagonal\,\, part}\right)_{\tilde{\ell}, \snu}
+\frac{1}{8\pi^2}(3m_0^2 + A_0^2)
{\bf Y}^{\dagger}_{{\nu}_{ik}} {\bf Y}_{{\nu}_{kj}} \log\frac{M_k}{M_{GUT}}\ .
\eea
The off-diagonal terms in  $m^2_{\tilde{\ell}, \snu}$  manifest
themselves in processes like $\mu \rightarrow e \gamma$ or 
$\tau \rightarrow \mu \gamma$, that could be observed in the
near future \footnote{The upper bounds on 
the off-diagonal entries of ${\bf Y}^{\dagger}_{{\nu}_{ik}}
{\bf Y}_{{\nu}_{kj}} \log\frac{M_X}{M_k}$
in the scenario of mSUGRA with the MSSM+3$\nu_R$s 
below the GUT scale apply for a wide class of models, since one 
does not expect cancellations among
the different terms in the RGEs, or with the off-diagonal elements
of the tree-level slepton mass matrix.}. 
In addition to this,  at tree level the 
three sneutrino masses are degenerate. 
However, radiative corrections induce a non-universality among
the masses that could perhaps be measured experimentally
\cite{Baer:2001cb}.  All these measurements could be used to disentangle
some information about the neutrino 
Yukawa  matrix and the right handed masses
from radiative corrections. 
See \cite{Ellis:2002fe} for a recent
analysis of $\ell_j \rightarrow \ell_i \gamma$ in this
approach. 

For leptogenesis we are particularly interested in the
phases of ${\bf Y}_{\nu}$, that are in turn related to the 
phases in the left-handed slepton mass matrix. It
is then an important issue to measure the phases in 
$m^2_{\tilde{\ell}, \snu}$. The electric dipole moments (EDMs)
of the electron and muon are CP violating
observables that could provide information about these phases.
However, this CP violation is flavour conserving
and could come from another flavour conserving sector of the
theory, like
the charginos and neutralinos, instead of the Yukawa couplings.
Furthermore, the EDMs care about the relative phases between
the charged leptons and the sleptons, of which the see-saw only 
induces one. So, the contribution induced by the see-saw would
be suppressed by small angles and Yukawa couplings \cite{RS}. Therefore,
to constrain the phases in the Yukawa couplings with the EDMs,
one has to make certain assumptions about the soft SUSY breaking
lagrangian.
A more detailed discussion  of
obtaining information about the complete 
theory  from low energy data  can be
found in \cite{Davidson:2001zk}.

The $V_L,D_Y$ low-energy parametrization has several advantages.
If we treat the 9 parameters of the neutrino mass matrix as 
``known'', there are 9 remaining unknown variables in
the seesaw: three phases and six
real numbers. Possible parametrizations of
these unknowns are $D_Y$ and $V_L$, $D_{\cal M}$ and the orthogonal 
complex matrix $R$ \cite{Casas:2001sr}, or as in \cite{Ellis:2002fe}.
The angles and phases of $V_L$ 
are related in a simple way to the lepton flavour violating slepton
mass matrix entries. These off-diagonal (in the charged
lepton mass eigenstate basis)  entries  are currently constrained
and could possibly be determined by radiative lepton
decays $\ell_j \rightarrow \ell_i \gamma$. The eigenvalues
of $D_Y$ are more difficult to determine experimentally. However,
we do measure the Yukawa matrix eigenvalues for the quarks and
charged leptons, so we can make theoretical guesses of the
$Y_{\nu}$  eigenvalues with more confidence than $e.g.$
guessing the $\nu_R$ Majorana masses.

It is convenient for our leptogenesis
analysis to parametrize the
sneutrino mass matrix with $D_Y$ and
$V_L$. It would be more correct to
express the lepton asymmetry in terms of
the magnitude and phases of slepton mass matrix
elements \footnote{We will follow this approach
in a subsequent publication \cite{DIP}.}.
Alternatively, there is an intermediate  parametrization,
which can be useful for analytic estimates.
The parameters we use,  $V_L$ and $D_Y$,  determine 
${\bf Y^{\dagger}_{\nu}} {\bf Y_\nu}$ rather than
${\bf Y}^{\dagger}_{{\nu}_{ik}} 
{\bf Y}_{{\nu}_{kj}} \log\frac{M_k}{M_{GUT}}$,
which is the expression that appears at leading log. 
It is easy, though,
to relate $V_L$ and $D_Y$  to ${\bf Y}^{\dagger}_{{\nu}_{ik}} 
{\bf Y}_{{\nu}_{kj}} \log\frac{M_k}{M_{GUT}}$. Noting that
\bea 
{\bf Y^{\dagger}_{\nu}}_{ik} \log{\frac{M_k}{M_{GUT}}} {\bf Y}_{{\nu}_{kj}}
&=&({\bf \widetilde Y^{\dagger}_{\nu}}  {\bf \widetilde Y_\nu})_{ij}=
({\widetilde V_L}^+ {\widetilde D_Y}^2 \widetilde V_L)_{ij} 
\\ \nonumber
\frac{{\cal M}_{\nu_{ij}}}{\langle H_u^0\rangle ^2} &=& 
{\bf \widetilde Y}_{{\nu}_{ik}}^T \frac{1}{{\widetilde M}_k} 
{\bf \widetilde Y}_{{\nu}_{kj}},
\label{Vtilde}
\eea
where 
${\bf \widetilde Y}_{ki}={\bf Y}_{ki}\sqrt{ \log{\frac{M_k}{M_{GUT}}}}$ and
${\widetilde M}_k=M_k  \log{\frac{M_k}{M_{GUT}}}$, it is possible to
rewrite eq.(\ref{see-saw2}) but using tilded parameters. So,
one could parametrize the see-saw mechanism with the neutrino mass matrix,
${\cal M}_{\nu}$, and ${\widetilde V}_L$, ${\widetilde D}_Y$, 
that are directly related to the leading-log
approximate solution of the left-handed slepton RGEs. Also,
from the definitions, it is straightforward to relate 
$V_L$ and $D_Y$ with their tilded-counterparts.
However, since
SUSY has not yet been discovered, 
we use $V_L$ and $D_Y$, with the knowledge
that we can calculate $[m_{\tilde{\nu}}^2]$ from
these parameters. This choice will be important when
we discuss phase overlaps.

We turn now to expressing the CP asymmetry in terms
of neutrino masses, the MNS matrix,  and other unknown parameters
 encoded in $D_Y$ and $V_L$.
We can make an analytic approximation 
indicating the dependence of   the  CP asymmetry $\epsilon$
 on our low energy parameters.
To derive these estimates, we first assume
$M_3 \gg M_1$ and $y_1 \ll y_2, y_3$. Then
we assume that $[{\cal M}^{-1 \dagger}{\cal M}^{-1}]_{11} $ is the largest
element of  ${\cal M}^{-1 \dagger} {\cal M}^{-1}$, in the basis where
 $Y_{\nu}$ is diagonal. 
As we will see, this is usually reasonable.

If a matrix $\Lambda$   has a zero eigenvalue, then 
the remaining two eigenvalues are
\beq
\lambda_1, \lambda_2 = 
  \frac{1}{2} \left\{ {\rm Tr}{ \Lambda} \pm
\sqrt{ \left({\rm Tr}{\Lambda}\right)^2 - 4  
\left({\Lambda}_{11}{\rm tr}{ \Lambda}
+{\rm det}{ \Lambda} - |{ \Lambda}_{12}|^2
  - |{\Lambda}_{13}|^2 \right)} \right\} ~~,
\label{1/M12}
\eeq
where Tr is the trace of the 3-d matrix,
and tr and det are defined on the 2-3 subspace.
In the limit where $M_3 \rightarrow \infty$, 
this formula can be applied to
the hermitian matrix $ {\cal M}^{-1 \dagger} {\cal M}^{-1}$:
\beq
{\cal M}^{-1 \dagger}{\cal M}^{-1} 
= \frac{ D_Y^{-1} V_L  {\cal M}_\nu^{\dagger} V_L^TD_Y^{-2} V_L^*  {\cal M}_\nu V_L^{\dagger} D_Y^{-1} }{ \langle H_u^0 \rangle^4}
\equiv \frac{ \Lambda}{y_1^4} ~~.
\eeq
To obtain simple expressions, we would like to
expand eq. (\ref{1/M12})  in small dimensionless
parameters. So to avoid confusion, we scale
a factor $y_1^4$ out of ${\cal M}^{-1} {\cal M}$.

The largest eigenvalue of $\Lambda$
will be of order $ m_{\nu}^2$, as can be seen by
defining $
\eta \equiv y_1  D_Y^{-1} = diag \left\{ \eta_1, \eta_2, \eta_3 \right\}$ 
and
\beq
\Delta = V_L^*  {\cal M}_\nu V^{\dagger}_L 
= V_L^* U^* D_{ {\cal M}_\nu} U^{\dagger} V_L^{\dagger}
\equiv W^* D_{ {\cal M}_\nu} W^{\dagger} ~~,
\label{Delta}
\eeq
which  gives  
\bea
\Lambda & =& 
  \frac{ \eta \Delta^{\dagger} \eta^2  \Delta \eta  }
  { \langle H_u^0 \rangle^4}~~.
\label{omega}
\eea
We take $y_3 = 1$. The matrix $W=V_L U$ is 
the rotation from the basis where the $\nu_L$ masses
are diagonal to the basis where the neutrino Yukawa  matrix
${\bf Y^{\dagger}_{\nu}} {\bf Y_\nu}$ is diagonal.

The dominant contributions to the matrix elements of $\Lambda$
can be calculated as an expansion in $\eta_2$ and
$\eta_3$.
Only $  \Lambda_{11}$ is zeroth order in
$\eta_2$ and $\eta_3$, so generically
 $ \Lambda_{11} \gg \Lambda_{22}, \Lambda_{33}$.
 Then  from eq.
(\ref{1/M12}), the lightest RH neutrino has mass
\beq
|M_1|^2 \simeq \frac{y_1^4 }{ \Lambda_{11}} ~~,
\eeq
and the associated eigenvector will be
\beq
\label{eigenvec}
\vec{v}_1 \simeq \left(\begin{array}{c}
\Lambda_{11} \\
\Lambda_{21} \\
\Lambda_{31}
\end{array} \right)  \times \frac{1}{\Lambda_{11}}
= \left(\begin{array}{c}
\Delta_{11}^* \\
\eta_2  \Delta^*_{12} \\
\eta_3  \Delta^*_{13}
\end{array} \right)  \times \frac{1}{\Delta_{11}^*} ~~.
\eeq

We can use this eigenvector to evaluate eq. (\ref{eps1}),
and find
\beq
\epsilon \simeq 
- \frac{3 y_1^2 \Lambda_{11}^2}{8 \pi [\Lambda D_Y^2 \Lambda]_{11} }
{\rm Im} \left\{ 
\frac{[\Lambda D_Y \Delta^\dagger D_Y \Lambda^T]_{11}}
{[\Lambda \eta  \Delta^\dagger \eta \Lambda^T]_{11}}\right\}
=
\frac{3 y_1^2}{8 \pi \sum_j |W_{1j}|^2  m_{\nu_j}^2} 
 {\rm Im} \left\{ \frac{ \sum_k W_{1k}^{2} m_{\nu_k}^3 }
{ \sum_n W_{1n}^{2} m_{\nu_n} } \right\} ~~,
\label{epsapprox}
\eeq
where we have dropped terms of order  $\eta_2$ and $\eta_3$,
and 
recall that $W$ 
is the rotation from the basis where the $\nu_L$ masses are
diagonal to the basis
where ${\bf Y^{\dagger}_{\nu}} {\bf Y_\nu}$ is diagonal.

It is important to notice that the CP asymmetry depends
only on the first row of the matrix $W$, that in turn depends only on
the first row of $V_L$. In the parametrization that we have chosen for
$V_L$, eq.(\ref{VLdef}), the first row depends on two angles and two
phases. Therefore, at the end of the day, the CP asymmetry 
depends on the neutrino mass matrix and
five unknown parameters: $y_1$, two angles and two phases.
Note that for generic $\Delta$, the order of magnitude of
$\epsilon$ is fixed by $y_1^2$. For the GUT-inspired 
value $y_1 \simeq m_u/m_t \sim 10^{-4}$,  $\epsilon \sim 10^{-9}$
unless there is some amplification in
$\mathrm{Im} \{ [\Delta \Delta^{\dagger} \Delta ]_{11} \Delta_{11}^* \}/
( [\Delta \Delta^{\dagger}]_{11} | \Delta_{11}|^2 )$.

\section{Phases for leptogenesis}
\label{5}

From the previous discussion, we find that in the parametrization
we have chosen, the CP asymmetry depends on five phases,
namely the phases in the MNS matrix, $\delta$, $\phi$ and
$\phi'$, and the phases in the first row of the $V_L$ matrix, 
$\varphi_{12}$ and $\varphi_{13}$. 
In this section we would like to study
the relative importance of these phases on the CP asymmetry,
and whether any of them could be considered as the ``leptogenesis
phase'', i.e. the phase that is fully responsible of the
CP asymmetry.  

To this end, we  first  introduce  a definition 
of ``overlap'' between the ``leptogenesis phase''
and the low energy phases. At the end of the section,
we will discuss issues raised by our definition. 

We define the  contribution to the CP asymmetry from a phase
$\alpha$ (this is {\it not} quite what we call
a phase overlap)   such that the total
CP asymmetry is the sum of the different contributions:
\bea
\label{definition}
\epsilon = \epsilon_{\delta}+\epsilon_{\phi}
+\epsilon_{\phi'}+\epsilon_{\varphi_{12}}+\epsilon_{\varphi_{13}} ~~.
\eea

To obtain a decomposition of the CP asymmetry in this way,
and give a more precise definition of the different contributions,
we Fourier expand the CP asymmetry:
\bea
\label{fourier}
\epsilon   =   \sum_{j,k,l,m,n} A_{j k l m n} \sin(j \delta 
+ k \phi  + l \phi' + m \varphi_{12} + n \varphi_{13}) ~~.
\eea 

This summation can be split in 
\bea
\label{terms}
\epsilon   =   
\sum_{\alpha} C_{\alpha} + 
\sum_{\alpha < \beta} C_{\alpha \beta}+
\sum_{\alpha < \beta < \gamma} C_{\alpha \beta \gamma}+
\sum_{\alpha < \beta < \gamma < \rho} C_{\alpha \beta \gamma \rho}+
\sum_{\alpha < \beta < \gamma < \rho < \sigma} C_{\alpha \beta \gamma \rho \sigma} ~~,
\eea
with $\{\alpha, \beta, \gamma, \rho, \sigma \}$ elements of
the ordered set 
$\{\delta, \phi, \phi', \varphi_{12}, \varphi_{13} \}$. 
Note  that the subindices of the C's  are ordered, so
$\delta <  \phi <  \phi' <  \varphi_{12} <  \varphi_{13}$.
 (to avoid double counting, only $C_{\delta \phi}$
exists, and $C_{\phi \delta}$ does not.)
Some of the terms in the summation are:
\bea
\label{Cdelta}
C_{\delta}=\sum_{j \neq 0} A_{j 0 0 0 0} \sin( j \delta)
\eea
\bea
\label{Cdeltaphi}
C_{\delta \phi}=\sum_{j \neq 0, k\neq 0} A_{j k 0 0 0}
\sin(j \delta + k \phi)
\eea
\bea
\label{Cdeltaphiphip}
C_{\delta \phi \phi'}=\sum_{j \neq 0, k\neq 0, l \neq 0} A_{j k l 0 0}
\sin(j \delta + k \phi + l \phi')
\eea
and so on.

We can now rewrite the summation eq.(\ref{terms}) in a way that
resembles eq.(\ref{definition}).
It is clear that $C_{\delta}$ is a contribution from $\delta$ to the
CP asymmetry, so it must be one of the terms in $\epsilon_{\delta}$. 
On the other hand, $C_{\delta \phi}$ is a contribution
from $\delta$, but also from $\phi$, and it is not possible to 
conclude whether it is a contribution from $\delta$ or from $\phi$.
So, we will say that $C_{\delta \phi}$ contributes
in $C_{\delta \phi}/2$ to $\epsilon_{\delta}$ and in 
$C_{\delta \phi}/2$ to $\epsilon_{\phi}$. This rationale can
be applied to the rest of the terms in the expansion eq.(\ref{terms})
to finally obtain 
\bea
\label{epsdelta}
\epsilon_{\delta}   =   
C_{\delta} + 
\frac{1}{2}\sum_{\beta} C_{\delta \beta}+
\frac{1}{3}\sum_{\beta < \gamma} C_{\delta \beta \gamma}+
\frac{1}{4}\sum_{\beta < \gamma < \rho} C_{\delta \beta \gamma \rho}+
\frac{1}{5}\sum_{\beta < \gamma < \rho < \sigma} C_{\delta \beta \gamma \rho \sigma} ~~,
\eea
and similarly for $\epsilon_{\phi}$, $\epsilon_{\phi'}$,
$\epsilon_{\varphi_{12}}$ and $\epsilon_{\varphi_{13}}$. It can be checked
that with this decomposition, eq.(\ref{definition}) holds.

In this analysis, we are  only concerned
with the relative contributions of the different phases
to the CP asymmetry, and not with the overall magnitude of the 
CP asymmetry itself (that is  essentially determined by the unknown
parameter $y_1$). So, we  normalize the different contributions
to 1, and define ``phase overlap'' as the
normalized contribution from a  phase to the CP asymmetry.
This quantity measures the relative importance of that phase for
the CP asymmetry compared to the rest of the phases.
For instance, the ``overlap'' of $\delta$ with the leptogenesis phase is:
\beq
O_{\delta} = \frac{|\epsilon_{\delta}|}
{\sqrt{ \sum_\alpha \epsilon_\alpha^2}} ~~,
\label{overlap}
\eeq
which satisfies $\sum O_\alpha^2 = 1$ and $0 \leq O_\alpha \leq 1$.
Had we chosen a linear normalization, i.e. 
$O_\alpha   = \epsilon_\alpha/ \epsilon$, in some regions
of the parameter space  $|O_\alpha|$ could be larger than one,
due to cancellations among the different $ \epsilon_\alpha$'s.
So, we prefer to use a quadratic normalization, that satisfies 
$0 \leq O_\alpha \leq 1$ to better represent the fractional
contribution of the phase $\alpha$ to $\epsilon$.

The overlap  defined in eq.(\ref{overlap}) measures
the importance of $\delta$ for $\epsilon$, provided that
the phases in the expansion are independent and
``orthogonal''. In any
parametrization of the seesaw, six phases are required,
so independence is automatic.
 The importance of ``orthogonality'' can be understood by
analogy with linear algebra, where a vector can be uniquely
decomposed in its components along a given orthonormal basis.
Similarly, the phases of our parametrization
must be ``orthogonal'', as well as independent, to
have a unique definition of the fraction of $\epsilon$ due to
$\delta$. However, a mathematical
definition of ``orthogonality'' is difficult, perhaps impossible,
because we do not have an inner product between phases. 
So we opt for a physical notion: we assume as ``orthogonal''
the so-called ``physical'' phases, 
the phases that could be measured at low energy --- in practice,
or in principle in the best of all physicists worlds. 
Notice that the choice of  low energy phases
 is important --- had we  parametrized with  the phases
of $U$ and $W$, then from eq. (\ref{epsapprox}), $\epsilon$ 
depends only on the phases of $W_{1i}$ and is independent
of the MNS phases. Similarly, if the seesaw is
parametrized using $U$ and the complex orthogonal matrix
$R$, The MNS matrix cancels out of the equation
for $\epsilon$, and the $\delta$ dependence of $\epsilon$
is buried in $R$.

The measurable phases of the slepton sector are those of
the slepton mass matrices, so we should expand $\epsilon$ on $\delta$,
$\phi'$, $\phi$, and the phases of $[m^2_{\tilde L}]$. 
However, we find it convenient to parametrize
the seesaw in terms of $V_L$ and
$D_Y$, rather than $[m^2_{\tilde L}]$, as discussed
in section \ref{4}. 
The  $[m^2_{\tilde L}]$ are therefore functions of the real 
angles of $V_L$, as well as the
phases of $V_L$, so $\varphi_{12}$ and $\varphi_{13}$
are not quite the correct physical phases. 
We expect this choice to have little effect on the
relative importance of $\delta$ and $\phi'$ for
leptogenesis. $V_L$ is closely related to 
the slepton mass matrix, and in the limit that
the real angles  in $V_L$ are small, the phases of 
$[m^2_{\tilde L}]$ are those of $V_L$ (or $\tilde{V}_L$) in
leading log. 


Finally, this definition of overlap
is {\it statistical }. $O_\delta$ gives the probable importance
of $\delta$ for $\epsilon$, assuming all phases are unknown and
$O(1)$. But if, for instance, $\delta = 0$, 
$O_\delta$ will be non-zero, because parts of
the crossed terms $\{ C_{\delta \alpha}, 
C_{\delta, \alpha, \beta}, ... \}$ are included
in $O_\delta$, and these terms could be non-zero due
to the other phases.


To avoid possible confusion, observe that
our notion of ``leptogenesis phase'' differs from
the one introduced in \cite{Hamaguchi:2001gw}, who write
$\epsilon = \epsilon_{max}  \delta_{eff}$ where
$\epsilon_{max}$ is the upper bound on $\epsilon$.
The asymmetry is the imaginary part of
a complex number  $\epsilon \equiv {\rm Im} \{ |\epsilon_c|e^{i \rho} \} =
 |\epsilon_c| \sin \rho $, so we interpret
the ``leptogenesis phase'' to be $\rho$.

The definition, eq. (\ref{overlap}), of the fraction
of $\epsilon$ that is due to $\delta$, depends on
eight unknowns: three real angles and
the five phases. We assume that the real
angles could be measured, so we present results
for different fixed values of the angles. We take
random values of the phases, linearly distributed
between 0 and $2 \pi$, and make
scatter plots of the overlaps $\{ O_\alpha \}$.
If most of the points are distributed at 
$|O_\alpha|^2 > .3$, we conclude that the phase
$\alpha$ contributes significantly to $\epsilon$
($\alpha$ here represents any phase among $\{ \delta, \phi', \phi,
\varphi_{12}, \varphi_{13} \}$). Notice however
that this is a statistical statement, based
on choosing all the phases randomly and large.

\section{The case $V_L=1$}
\label{6}

In this section we particularize the previous study to the 
case $V_L=1$. In this case, there is no 
flavour or CP violation induced radiatively 
by right-handed neutrinos in the slepton mass matrices.
Since we have fixed the $V_L$ matrix, the number of unknown parameters
is reduced, and the CP asymmetry depends just on the neutrino mass matrix
and $y_1$, the lightest eigenvalue of 
${\bf Y^{\dagger}_{\nu}} {\bf Y_\nu}$.

When $V_L=1$, only the phases in the MNS matrix ($\delta$, $\phi$ 
and $\phi'$) are relevant, hence the analysis of the previous section
is greatly simplified. The CP asymmetry can be written as the
sum of the contributions from the phases $\delta$, $\phi$ 
and $\phi'$,
\bea
\label{definition1}
\epsilon = \epsilon_{\delta}+\epsilon_{\phi}+\epsilon_{\phi'}  ~~.
\eea
As in the previous section, we Fourier expand $\epsilon$, 
yielding
\bea
\label{fourier1}
\epsilon   =   \sum_{j,k,l} A_{j k l} \sin(j \delta+k \phi+l \phi')
= C_{\delta}+C_{\phi}+C_{\phi'}+
 C_{\delta \phi} +C_{\delta \phi'} +C_{\phi \phi'} +
C_{\delta \phi \phi'} ~~,
\eea
with $C_{\alpha}$, $C_{\alpha \beta}$ and $C_{\delta \phi \phi'}$ as
in eqs.(\ref{Cdelta})-(\ref{Cdeltaphiphip}). Then, the contribution
from the phases $\delta$, $\phi$ and $\phi'$ to the CP asymmetry
read 
\bea
\label{epsdelta1a}
\epsilon_{\delta}   &= &  
C_{\delta} + 
\frac{1}{2} (C_{\delta \phi} +C_{\delta \phi'})+
\frac{1}{3} C_{\delta \phi \phi'} \nonumber \\
\epsilon_{\phi}  & =  & 
C_{\phi} + 
\frac{1}{2} (C_{\delta \phi} +C_{\phi \phi'})+
\frac{1}{3} C_{\delta \phi \phi'} \\
\epsilon_{\phi'}  & = &  
C_{\phi'} + 
\frac{1}{2} (C_{\delta \phi'} +C_{\phi \phi'})+
\frac{1}{3} C_{\delta \phi \phi'}  ~~. \nonumber
\eea

As before, and since we are only concerned with the relative contributions
from the different phases to $\epsilon$ and not with the overall
magnitude, we define the ``phase overlaps'' as:
\bea
\label{eps_norm1}
O_{\alpha}=\frac{|\epsilon_{\alpha}|}
{\sqrt{\epsilon_{\delta}^2+\epsilon_{\phi}^2+\epsilon_{\phi'}^2}},
\eea
where $\alpha={\delta, \phi, \phi'}$. With these definitions,
the following identity holds:
\bea
\label{triangle}
(O_{\delta})^2+(O_{\phi})^2+
(O_{\phi'})^2=1 ~~.
\eea

In Figure 1 we show the numerical results for different CHOOZ
angles. We show the results for the LAMSW solution
to the solar neutrino problem and the mass hierarchy 
$m_{\nu_1}:m_{\nu_2}:m_{\nu_3} = 10^{-2}:0.1:1$. Each point
corresponds to a random value of the phases $\delta$, $\phi$
and $\phi'$ between 0 and $2 \pi$. In view of eq.(\ref{triangle}),
we find convenient to present the results on a triangular plot,
where the distance to the sides of the triangle corresponds to the
``phase overlaps'' squared, defined in eq.(\ref{eps_norm1}) 
(see upper left plot). 

When the CHOOZ angle is close to the experimental bound (upper right plot)
over most of the parameter space the relevant phases are 
$\delta$ and $\phi'$, and their contributions are approximately
equal. In this case, the phase $\phi$ is essentially irrelevant,
except for a few points that correspond to 
$2 \delta - \phi' \simeq 0, \pi$. On the other hand, when 
the CHOOZ angle is moderately small (lower left plot),
we find points scattered over the
whole triangle: the three phases are relevant in this case. One
can also see from the figure that the points 
seem to follow a circular pattern. 
We will come back to this issue later on. Finally, when the CHOOZ
angle is very small (lower right plot), 
the relevant phases are $\phi$ and $\phi'$, except
for the points for which $\phi-\phi' \simeq 0, \pi$, where
the phase $\delta$ becomes relevant. 
A neutrino factory is expected to be sensitive to
$\sin \theta_{13} \gsim 10^{-4}$ \cite{Cervera:2000kp} and
to be able to see CP violation for phases of order 1
if $\sin \theta_{13} \gsim .01$ \cite{Romanino:1999zq}.

\begin{figure}
\centerline{\hbox{
\psfrag{Od}{\Large $O^2_{\delta}$}
\psfrag{Op}{\Large $O^2_{\phi}$}
\psfrag{Opp}{\Large $O^2_{\phi'}$}
\psfig{figure=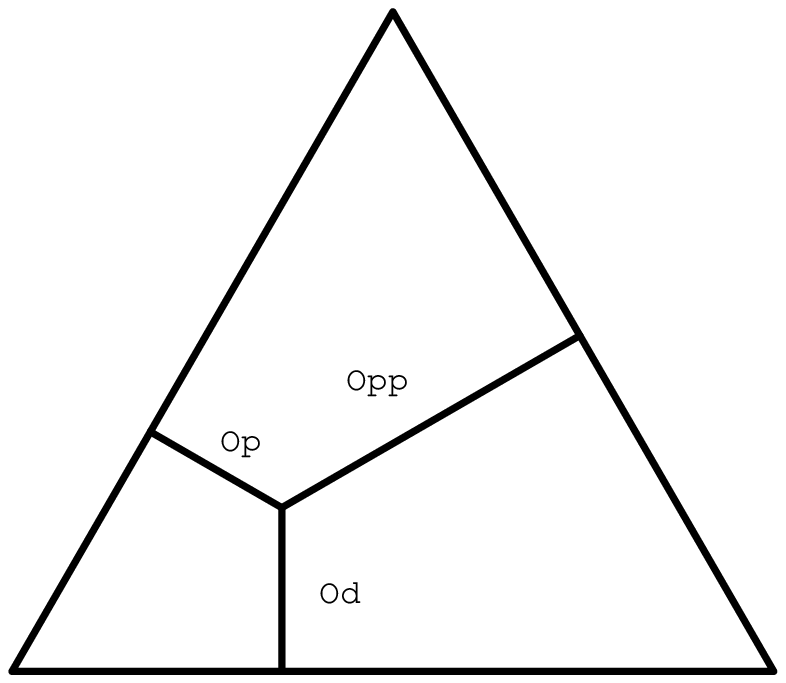,height=10.4cm,width=11.4cm,bbllx=1.cm,%
bblly=0.1cm,bburx=12.cm,bbury=9.1cm}
\psfig{figure=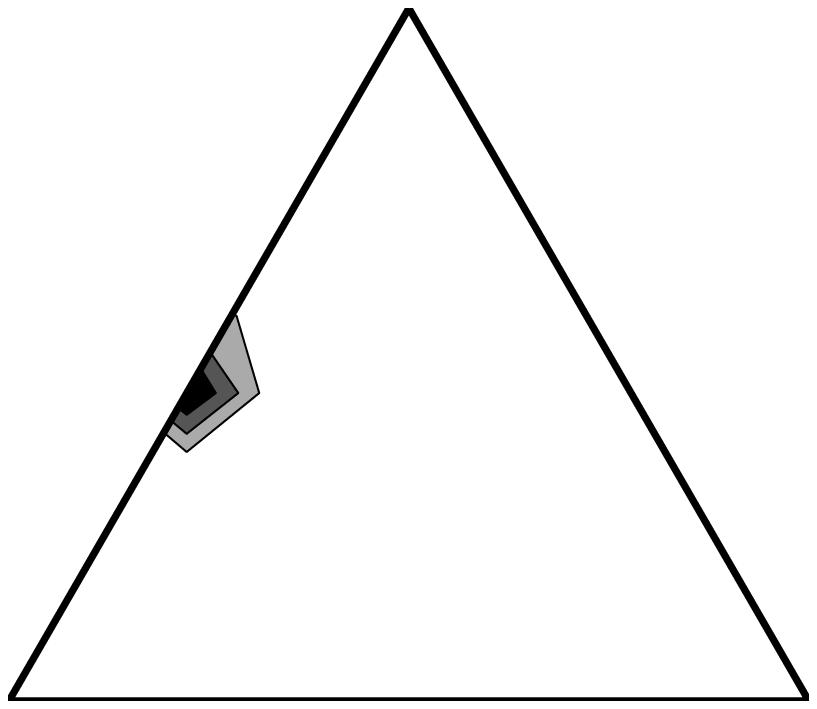,height=10cm,width=11cm,bbllx=3.cm,%
bblly=0.cm,bburx=14.cm,bbury=9.cm}}}
\centerline{\hbox{
\psfig{figure=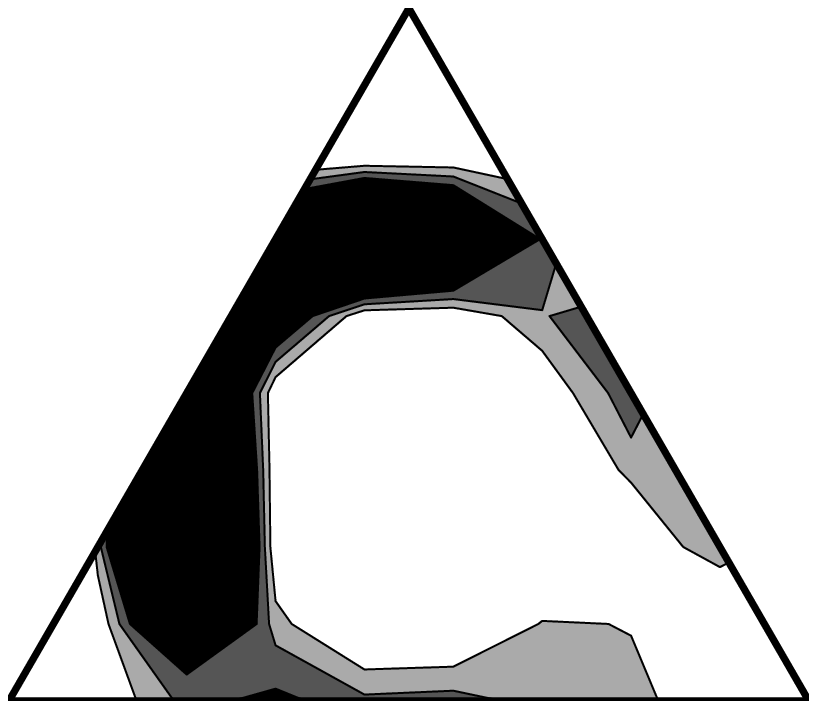,height=10cm,width=11cm,bbllx=1.cm,%
bblly=0.cm,bburx=12.cm,bbury=9.cm}
\psfig{figure=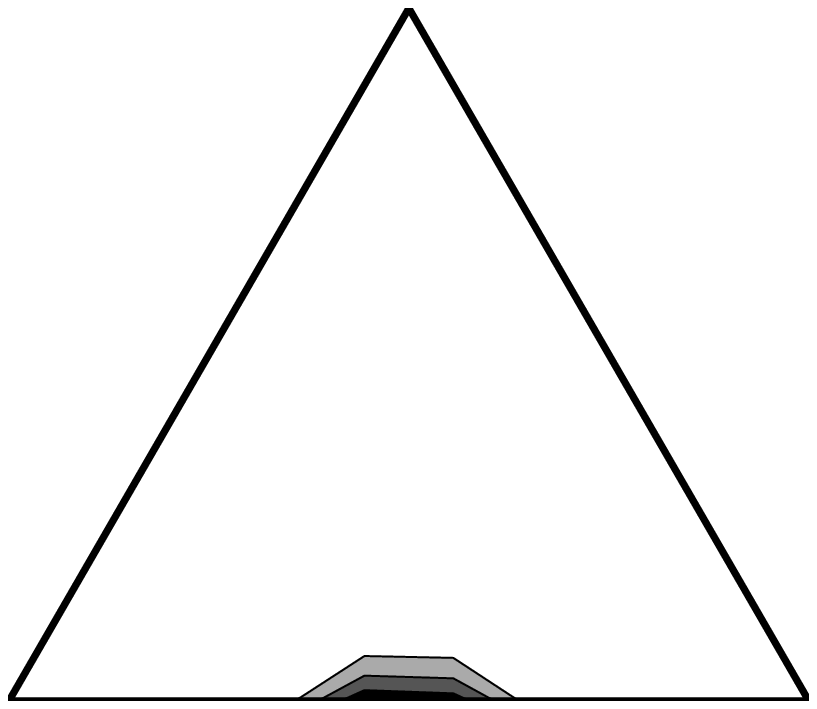,height=10cm,width=11cm,bbllx=3.cm,%
bblly=0.cm,bburx=14.cm,bbury=9.cm}}}
\caption
{\footnotesize 
``Phase overlaps'' for the case $V_L=1$, i.e. 
${\bf Y^{\dagger}_{\nu}} {\bf Y_\nu}$ diagonal. The upper left
plot indicates the meaning of the distances to the different
sides of the triangle. The rest of the triangles show
density plots of the ``phase overlaps'' for random values
of the phases and different CHOOZ angles: 0.1 (upper right),
0.01 (lower left) and 0.001 (lower right). The darkest regions
correspond to the largest density of points.
}
\label{fig1}
\end{figure}

These plots can be understood analytically using the approximation
for the CP asymmetry, eq.(\ref{epsapprox}). 
When $V_L=1$, the CP asymmetry has a
fairly simple expression in terms of low energy neutrino 
data and $y_1$, the lightest 
eigenvalue of ${\bf Y^{\dagger}_{\nu}} {\bf Y_\nu}$:
\bea
\label{eps_analytic1}
\epsilon \simeq \frac{3 y_1^2}{8 \pi D} ~
{\rm Im} \left\{\frac{
m_{\nu_1}^3 ~c_{13}^2 c_{12}^2 ~e^{ -i \phi}+
m_{\nu_2}^3 ~c_{13}^2 s_{12}^2~ e^{ -i \phi'}+
m_{\nu_3}^3 ~s_{13}^2 ~ e^{-2 i \delta}}
{m_{\nu_1}~ c_{13}^2 c_{12}^2 ~e^{ -i \phi}+
m_{\nu_2} ~c_{13}^2 s_{12}^2~ e^{ -i \phi'}+
m_{\nu_3} ~s_{13}^2 ~ e^{-2 i \delta}} \right\}  ~~,
\eea
where $D= m_{\nu_1}^2 ~c_{13}^2 c_{12}^2+
m_{\nu_2}^2~ c_{13}^2 s_{12}^2+ m_{\nu_3}^2 ~s_{13}^2$.
For the mass hierarchy and the ranges of CHOOZ angles 
that we are using, it turns out that $m_{\nu_2} \gg m_{\nu_1},~ 
m_{\nu_3} s_{13}^2$, so  we can expand the denominator in 
$\epsilon$. Approximating 
$\theta_{12}, \theta_{23} \sim \pi/4$, the result is:
\bea
\label{eps_analytic1_taylor}
\epsilon \simeq -\frac{3 y_1^2}{4 \pi } \left\{ 
\left(\frac{m_{\nu_3}}{m_{\nu_2}}\right)^3 2 s_{13}^2 \sin(2 \delta -  \phi')-
 \frac{m_{\nu_1}}{m_{\nu_2}} \sin (\phi-\phi')
 \right\}  ~~.
\eea
When the CHOOZ angle is much larger than 
$\sqrt{m_{\nu_1} m^2_{\nu_2}/m^3_{\nu_3}}$ the first term in
eq.(\ref{eps_analytic1_taylor}) dominates, unless $2 \delta -  \phi'$ is
close to 0 or $\pi$. This condition is satisfied in particular when
the lightest neutrino is very light, which is an
interesting physical possibility.
For the mass hierarchy that we have chosen, the condition above
reads $s_{13} \gg 0.01$, which is satisfied when
the CHOOZ angle is close the present experimental limit.
Recall that $\theta_{13} \gsim .01$ is required to detect
$\delta$ at a neutrino factory, so this limit would hold
if CP violation is observed at the neutrino factory.
The CP asymmetry in this case can be approximated by
\bea
\label{CHOOZ_large}
\epsilon \simeq -\frac{3 y_1^2}{2 \pi }  
\left(\frac{m_{\nu_3}}{m_{\nu_2}}\right)^3 s_{13}^2 \sin(2\delta - \phi'),
\eea
that does not depend on $\phi$; only on $\delta$ and $\phi'$.
Furthermore, the dependence is such that one cannot conclude
whether the "leptogenesis phase" is $\delta$ or $\phi'$. Instead,
in this limit, the "leptogenesis phase" 
is the combination $2 \delta - \phi'$. 
Comparing eq.(\ref{CHOOZ_large}) with the Fourier expansion, 
eq.(\ref{fourier1}), it follows that for most points, 
$\epsilon \simeq C_{\delta \phi'}$. Hence $\epsilon_{\phi} \simeq 0$,
$\epsilon_{\delta} \simeq \epsilon_{\phi'} \simeq C_{\delta \phi'}/2$.
($\epsilon_{\delta} \simeq \epsilon_{\phi'}$ is a consequence of
the fact that $\epsilon$ depends on 
a combination of $\delta$ and $\phi'$.)
Consequently, most points in the scatter plot, Fig. 1, upper right, 
are concentrated in the middle of the side corresponding to 
$O_{\phi} = 0$.

On the other hand, when the CHOOZ angle is very small, it is the
second term in eq.(\ref{eps_analytic1_taylor}) the one that dominates,
as long as 
$(\phi-\phi')$ is different from 0 or $\pi$. In this limit,
\bea
\label{CHOOZ_small}
\epsilon \simeq \frac{3 y_1^2}{4 \pi} 
 ~ \frac{m_{\nu_1}}{m_{\nu_2}} \sin (\phi-\phi') ~~.
\eea
The CP asymmetry only depends on $\phi$ and $\phi'$, and the
"leptogenesis phase" is $\phi-\phi'$. As before, comparing 
eq.(\ref{CHOOZ_small}) with the expansion eq.(\ref{fourier1}),
we conclude that for most points,
$\epsilon \simeq C_{\phi \phi'}$. Hence, $\epsilon_{\delta} \simeq 0$,
$\epsilon_{\phi} \simeq \epsilon_{\phi'} \simeq C_{\phi \phi'}/2$. 
In consequence, most points in Fig.1, lower right, are concentrated 
in the middle of the side corresponding to
$O_{\delta} = 0$. 

Finally, for  values of the CHOOZ angle $\sim 
\sqrt{m_{\nu_1} m_{\nu_2}^2/m_{\nu_3}^3}$,
 both terms in
eq.(\ref{eps_analytic1_taylor}) have to be taken into account. 
In this case, we cannot say that there is a single "leptogenesis phase":
both $\delta -2\phi$ and $\phi-\phi'$ are "leptogenesis phases".
Concerning the contributions from the phases
$\delta$, $\phi$ and $\phi'$ to the CP asymmetry, it is apparent from
eq.(\ref{eps_analytic1_taylor}) that in a generic point 
the three contributions are going to be comparable. To be precise:
\bea
\label{epsdelta1b}
\epsilon_{\delta}   &= &  
\frac{1}{2} C_{\delta \phi'} \nonumber \\
\epsilon_{\phi}  & =  & 
\frac{1}{2} C_{\phi \phi'} \\
\epsilon_{\phi'}  & = & 
\frac{1}{2} (C_{\delta \phi'} +C_{\phi \phi'})  \nonumber
\eea
where,
\bea
\label{epsdelta2}
C_{\delta \phi'} &\simeq&  -\frac{3 y_1^2}{2 \pi} 
~ \left(\frac{m_{\nu_3}}{m_{\nu_2}}\right)^3 s_{13}^2 
\sin(2 \delta -  \phi')  \nonumber\\
C_{\phi \phi'} &\simeq&  \frac{3 y_1^2}{4 \pi } 
~ \frac{m_{\nu_1}}{m_{\nu_2}} \sin (\phi-\phi')   ~~.
\nonumber
\eea
From these formulas, it is possible to understand the circular
pattern that appears in Fig. 1, lower left, 
changing from triangular coordinates
to Cartesian coordinates. We define the Cartesian axes setting the
origin at the lower left vertex of the triangle, and we denote
as $x$ ($y$) the horizontal (vertical) axis. The change of
variables read,
\bea
x&=& \frac{2}{\sqrt{3}} \left[ \frac{O_{\delta}^2}{2}+
O_{\phi}^2 \right] \nonumber \\
y&=&O_{\delta}^2  ~~,
\eea
and using 
that $\epsilon_{\phi'} \simeq \epsilon_{\delta}+\epsilon_{\phi}$, 
we obtain, after some algebra, 
$(x-\frac{1}{\sqrt{3}})^2+(y-\frac{1}{3})^2 \simeq \frac{1}{9}$, which 
is the equation of a circle centered in the barycentre of the triangle,
with radius $1/3$.

\section{The general case}
\label{7}

In the general case the number of unknown parameters involved is 
rather large (five phases, two angles in the $V_L$ matrix and
the CHOOZ angle), so the analysis is much more intricate since many
different limits arise. However, we will see that only a few limits
are distinct and physically interesting; 
the rest correspond to small regions
in the parameter space that could arise in
particular models, but that we will not consider, following the same
bottom-up spirit as in the rest of the paper. 

The different limits stem from the possible ways to expand 
the denominator in our approximate expression for the CP asymmetry
\bea
\label{epsilon-low2}
\epsilon & \simeq & 
\frac{3 y_1^2}{8 \pi \sum_n |W_{1n}|^2 m_{\nu_n}^2 } {\rm Im} \left\{ 
\frac{ \sum_i W_{1i}^{2} m_{\nu_i}^3 } {\sum_j W_{1j}^{2} m_{\nu_j} }
\right\}.
\eea 
The relevant elements in $W$ for the calculation are
\bea
W_{11}&\simeq&e^{-i \phi/2} \left[\frac{1}{\sqrt{2}}~c^L_{12} c^L_{13} +
e^{i \varphi_{12}} s^L_{12} c^L_{13}
\left(\frac{1}{2}+ \frac{1}{2} e^{i \delta} s_{13} \right)
-e^{i \varphi_{13}} s^L_{13}
\left(\frac{1}{2}- \frac{1}{2} e^{i \delta} s_{13} \right) \right] 
\nonumber \\
W_{12}&\simeq&e^{-i \phi'/2} \left[\frac{1}{\sqrt{2}}~c^L_{12} c^L_{13} -
e^{i \varphi_{12}} s^L_{12} c^L_{13}
\left(\frac{1}{2}- \frac{1}{2} e^{i \delta} s_{13} \right)
+e^{i \varphi_{13}} s^L_{13}
\left(\frac{1}{2}+ \frac{1}{2} e^{i \delta} s_{13} \right) \right]  ~~,
\label{W(49)}
\\
W_{13}&\simeq&e^{-i \delta} c^L_{12} c^L_{13} s_{13}  
- \frac{1}{\sqrt{2}}~ e^{i \varphi_{12}} s^L_{12} c^L_{13}-
\frac{1}{\sqrt{2}}~e^{i \varphi_{13}} s^L_{13} 
\nonumber
\eea
where we have approximated $\cos \theta_{13} \simeq 1$ and
we have assumed maximal solar and atmospheric mixings.
It is apparent from these
equations that different limits are going to arise depending on
the mixing angles in $V_L$ and the CHOOZ angle. 
We find then an interesting interplay
between leptogenesis and lepton flavour violation, induced by
radiative corrections through 
${\bf Y^{\dagger}_{\nu}} {\bf Y_\nu}= V_L^{\dagger} D_Y^2 V_L$. However,
from the parametrization we have chosen for $V_L$, eq.(\ref {VLdef}), 
one realizes that the off-diagonal elements
 of ${\bf Y^{\dagger}_{\nu}} {\bf Y_\nu}$ 
depend also on $\theta^L_{23}$, that does not play 
any role in the CP asymmetry 
generated in the decay of the lightest right-handed 
neutrino
\footnote{It is important, though,
for the computation of the CP asymmetry generated in the decay 
of the heavier right-handed neutrinos, that
could be relevant, or even dominant, in some scenarios (particularly
in scenarios with non-thermal creation of right handed neutrinos). 
Research along this lines would be certainly interesting, since in this
case lepton flavour violation could be intimately related with 
leptogenesis.}.

We obtain simple analytic expressions when the denominator
of eq. (\ref{epsilon-low2}) can be expanded in small
parameters. In the whole of
section \ref{7}, we assume  $|W_{12}| >|W_{11}|\sqrt{ m_{\nu_1}/ m_{\nu_2}} $,
so we can neglect $m_{\nu_1}$ terms in the denominator.
 When  $|W_{13}|^2 <|W_{12}|^2 m^2_{\nu_2}/ m^2_{\nu_3} $
which will be the case if
 $\theta^L_{12}$,  $\theta^L_{13}$, $\theta_{13} < .1$,
$\epsilon$ can be approximated by
\bea
\label{expansion_small}
\epsilon \simeq \frac{3 y_1^2}{8 \pi |W_{12}|^2} {\rm Im} \left[ 
\left(\frac{m_{\nu_3}}{m_{\nu_2}}\right)^3 \frac{W_{13}^2}{W_{12}^2}-
\left(\frac{m_{\nu_1}}{m_{\nu_2}}\right) \frac{W_{11}^2}{W_{12}^2} \right]~~.
\eea
On the other hand, when the mixing in $V_L$ is large, in the sense that
$\sin  \theta^L_{12}$ or $\sin  \theta^L_{13}$ is larger than 
$\sim 0.1$, then $ |W_{13}|^2 >|W_{12}|^2 m_{\nu_2}/ m_{\nu_3} $,
 and
the CP asymmetry reads
\bea
\label{expansion_large}
\epsilon \simeq - \frac{3 y_1^2}{8 \pi |W_{13}|^2}   
\left(\frac{m_{\nu_2}}{m_{\nu_3}}\right)
 {\rm Im} \left[ \frac{W_{12}^2}{W_{13}^2} \right]~~.
\eea

There is also an intermediate case, between these limits,
where  $ |W_{13}|^2 m_{\nu_3}/ m_{\nu_2} < |W_{12}|^2  < 
|W_{13}|^2 m^2_{\nu_3}/ m^2_{\nu_2} $. We do not discuss
this, because $m_{\nu_3}/ m_{\nu_2} \sim 10$ for the
hierarchical LMA solution we consider.

Let us analyze the two  cases separately.

\subsection{$|W_{11}| \sim |W_{12}| \gg |W_{13}| m_{\nu_3}/ m_{\nu_2} $}

The analysis for this case is parallel to the one we performed in
the previous section for the case $V_L=1$, where 
$|U_{11}| \sim |U_{12}| \gg |U_{13}| m_{\nu_3}/ m_{\nu_2} $. 
Using that $m_{\nu_2} \gg
m_{\nu_1}$, $m_{\nu_3} s_i s_j$, where $s_i$ is any of $s_{13}$, 
$s^L_{12}$, $s^L_{13}$, we can expand 
eq.(\ref{expansion_small}), keeping the leading order terms in
the expansion. The result is:
\bea
\label{theta12-13_small}
\epsilon &\simeq& -\frac{3 y_1^2}{4 \pi} \left\{ 
\left(\frac{m_{\nu_3}}{m_{\nu_2}}\right)^3  
\left[2 s^2_{13} \sin(2 \delta - \phi') - (s^L_{12})^2 
\sin (2 \varphi_{12}+\phi') -(s^L_{13})^2 \sin (2 \varphi_{13}+\phi') 
\right. \right. \nonumber \\
& & 
\left. 
-2 \sqrt{2} s_{13} s^L_{12} 
\sin(\delta - \varphi_{12}-\phi') -
2 \sqrt{2} s_{13} s^L_{13} \sin(\delta - \varphi_{13}-\phi') 
-2  s^L_{12} s^L_{13} \sin(\varphi_{12} + \varphi_{13}+\phi') \right]
\nonumber \\
& &
\left.
-\left(\frac{m_{\nu_1}}{m_{\nu_2}}\right) 
\sin(\phi-\phi')
\right\} ~~.
\eea
Obviously, in the limit $s^L_{12},s^L_{13} \rightarrow 0$ we recover 
eq.(\ref{eps_analytic1_taylor}). In the case $V_L=1$ we found different
limits, depending on the value of the CHOOZ angle. Now, the role 
of the CHOOZ angle is played by the angles $\theta^L_{12}$,
$\theta^L_{13}$ and the CHOOZ angle itself, and the results
are different depending on the values of these angles compared
with $\sqrt{m_{\nu_1} m^2_{\nu_2}/m^3_{\nu_3}}$. 

\vspace{0.3cm}
$\bullet$ When {\it any} of the angles $s_{13}$, 
$s^L_{12}$ or $s^L_{13}$ is much larger than 
$\sqrt{m_{\nu_1} m^2_{\nu_2}/m^3_{\nu_3}}$,
the term proportional to $(m_{\nu_3}/m_{\nu_2})^3$ 
in eq.(\ref{theta12-13_small}) dominates:
\bea
\label{m3-dominates}
\epsilon &\simeq& -\frac{3 y_1^2}{4 \pi} 
\left(\frac{m_{\nu_3}}{m_{\nu_2}}\right)^3 
\left[  2 s^2_{13} \sin(2 \delta - \phi') - (s^L_{12})^2 
\sin (2 \varphi_{12}+\phi') -(s^L_{13})^2 \sin (2 \varphi_{13}+\phi') 
\right.\\
& & 
\left. -2 \sqrt{2} s_{13} s^L_{12} 
\sin(\delta - \varphi_{12}-\phi') -
2 \sqrt{2} s_{13} s^L_{13} \sin(\delta - \varphi_{13}-\phi') 
-2  s^L_{12} s^L_{13} \sin(\varphi_{12} + \varphi_{13}+\phi')
  \right] ~~.
 \nonumber
\eea
We recall here that this limit corresponds to the case where the
lightest neutrino mass is very small. On the other hand, for the
mass hierarchy that we are using as reference to present our
numerical results, $m_{\nu_1}:m_{\nu_2}:m_{\nu_3} = 10^{-2}:0.1:1$,
this limit arises when any of the angles is much larger than $\sim 0.01$,
in particular, when the CHOOZ angle is close to the experimental 
bound and the relevant angles in $V_L$ are comparable to or smaller
than the CHOOZ angle. 
 
When the three angles are comparable in size, we see that there 
are three ``leptogenesis phases'': $2 \delta - \phi'$, 
$2 \varphi_{12}+\phi'$ and $2 \varphi_{13}+\phi'$ (the arguments of the
sines in the last three terms of eq.(\ref{m3-dominates})
are combinations of these). Notice that
in this limit $\phi'$ is an important phase for leptogenesis, although it cannot be regarded as the ``leptogenesis phase'', since the actual
``leptogenesis phases'' are combinations of $\phi'$ with other phases.
However, an indication for a non-vanishing $\phi'$, coming for example
from experiments on neutrinoless double beta decay, would provide 
an indication for leptogenesis.

If there are two angles that are comparable, while the third is much 
smaller than the others, then there 
are two ``leptogenesis phases''. To understand better the results
for this limit, we analyze in some detail the case 
$s_{13} \simeq s^L_{12} \gg s^L_{13}$. If $s^L_{23}$ is also
small, this case would produce small 
rates for $\mu \rightarrow e \gamma$, as can be checked
from eq.(\ref{VLdef}).
The results for the other possibilities, $s_{13} \simeq s^L_{13} 
\gg s^L_{12}$ and $s^L_{12} \simeq s^L_{13} \gg s_{13}$, 
can be easily deduced from this analysis, 
making the appropriate substitutions.
We have computed numerically the different contributions to
the CP asymmetry for the choice of 
angles $s_{13} = s^L_{12}=0.03$, $s^L_{13}=0$, the mass hierarchy
$m_{\nu_1}:m_{\nu_2}:m_{\nu_3} = 10^{-2}:0.1:1$ 
and assigning random values, between 0 and $ 2 \pi$, to the phases.
We obtain that for most of the parameter space, 
the only non-vanishing contributions to the CP asymmetry 
are $\epsilon_{\delta}$, $\epsilon_{\varphi_{12}}$
and $\epsilon_{\phi'}$ ($\varphi_{13}$ does not play any role, because
we have set $s^L_{13}$ to $0$ ). Since there are essentially only 
three contributions involved, a convenient way of
presenting the results is using a triangular plot. 
In Fig.2, left, we explain how to interpret the distances to the different
sides of the triangle, whereas in Fig.2, right, we show the
numerical results of the calculation. We find that in general
the three contributions are comparable, although the contribution 
from $\phi'$ is slightly larger than the other two. 
This can be understood from 
the analytical approximation, eq.(\ref{m3-dominates}), setting
$s^L_{13}=0$. The different contributions to the CP asymmetry are:
\bea
\label{epsdelta1c}
\epsilon_{\delta}   &= &  
\frac{1}{2} C_{\delta \phi'}+
\frac{1}{3} C_{\delta \phi' \varphi_{12}} 
\nonumber \\
\epsilon_{\phi'}  & =  & 
\frac{1}{2} (C_{\delta \phi'} +C_{\phi' \varphi_{12}})+
\frac{1}{3} C_{\delta  \phi' \varphi_{12}} 
\\
\epsilon_{\varphi_{12}}  & = & 
\frac{1}{2} C_{ \phi' \varphi_{12}}+
\frac{1}{3} C_{\delta  \phi' \varphi_{12}}  ~~, \nonumber
\eea
where,
\bea
C_{\delta \phi'} &\simeq&  -\frac{3 y_1^2}{2 \pi} 
~ \left(\frac{m_{\nu_3}}{m_{\nu_2}}\right)^3  s_{13}^2 
\sin(2 \delta -  \phi')  
\nonumber \\
C_{ \phi' \varphi_{12} } &\simeq&  \frac{3 y_1^2}{4 \pi } 
~ \left(\frac{m_{\nu_3}}{m_{\nu_2}}\right)^3 (s^L_{12})^2 
\sin (2 \varphi_{12}+\phi' )
\\
C_{\delta \phi' \varphi_{12}}  &\simeq&- \frac{3 y_1^2}{\sqrt{2} \pi } 
~ \left(\frac{m_{\nu_3}}{m_{\nu_2}}\right)^3  s_{13} s^L_{12} 
\sin(\delta - \varphi_{12}-\phi'),
\nonumber
\eea
that are in general comparable. 
\begin{figure}
\centerline{\hbox{
\psfrag{Op12}{\Large $O^2_{\varphi_{12}}$}
\psfrag{Od}{\Large $O^2_{\delta}$}
\psfrag{Opp}{\Large $O^2_{\phi'}$}
\psfig{figure=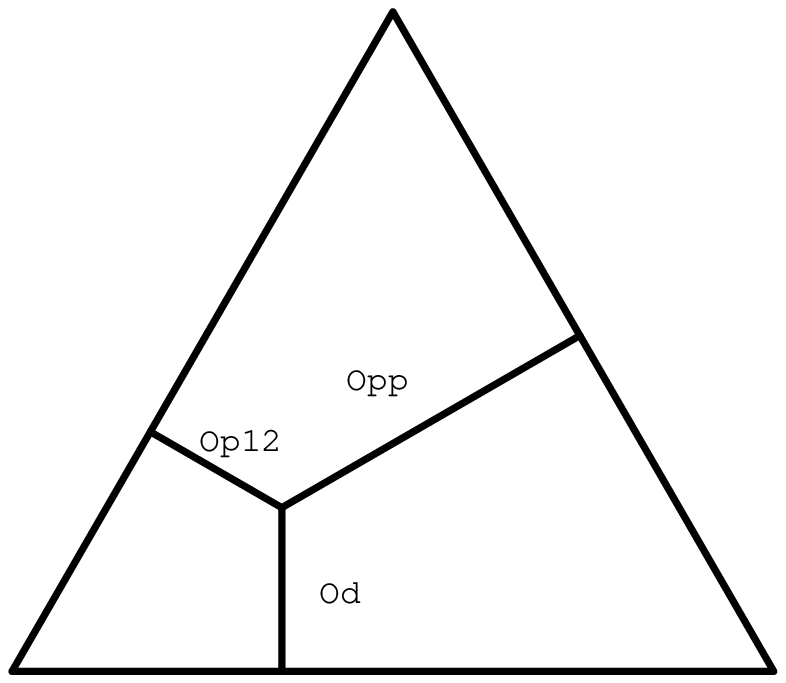,height=10.4cm,width=11.4cm,bbllx=1.cm,%
bblly=0.1cm,bburx=12.cm,bbury=9.1cm}
\psfig{figure=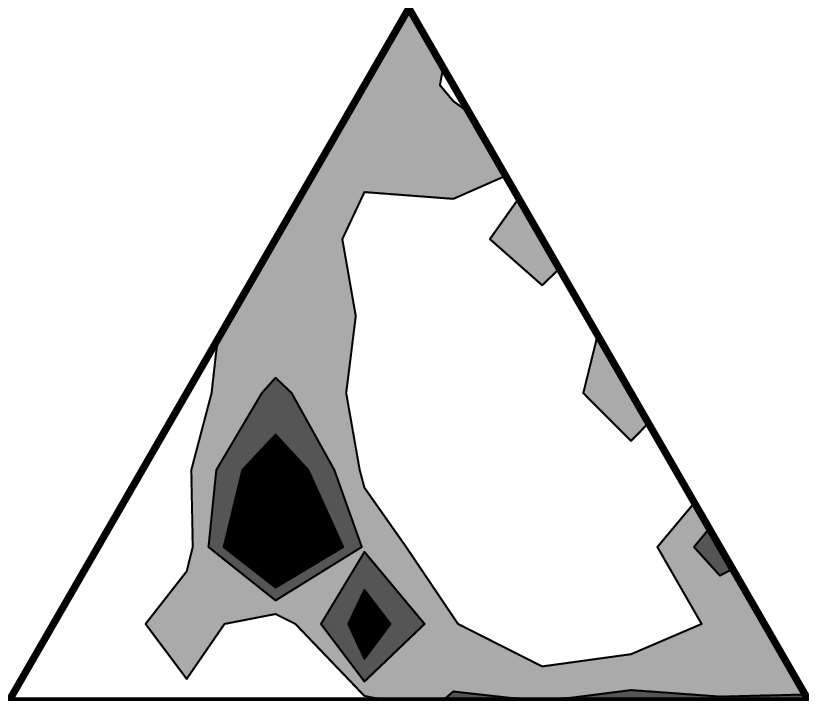,height=10cm,width=11cm,bbllx=3.cm,%
bblly=0.cm,bburx=14.cm,bbury=9.cm}}}
\caption
{\footnotesize The same as Fig.1, for the case where 
 $s_{13} = s^L_{12}=0.03$, $s^L_{13}=0$. The left plot indicates
how to interpret the distances to the different sides
of the triangle, and the right plot shows a density
plot of the ``phase overlaps'' for random values of
the phases. The darkest regions correspond to 
the largest density of points.
}
\label{fig2}
\end{figure}

Finally, if one of the angles dominates over the others, the conclusions are
very similar as for the case $V_L=1$, where the CP asymmetry received
contributions from $\phi'$ and $\delta$. Here, the role of $\delta$ is 
played by the phase corresponding to the angle that dominates 
($\delta$ for $s_{13}$, $\varphi_{12}$ for $s^L_{12}$, 
and $\varphi_{13}$ for $s^L_{13}$). In this case, 
$\epsilon \simeq C_{x \phi'}$, where $x$ is the relevant
angle among $\delta$, $\varphi_{12}$ and $\varphi_{13}$. On the 
other hand, the normalized contributions are $O_{x} 
\simeq O_{\phi'} \simeq 1/\sqrt{2}$, while they are
vanishing for the rest of the phases. For example,
if $s^L_{12} \gg s_{13}, s^L_{13}$, then 
$\epsilon \simeq C_{\phi' \varphi_{12}}$ and 
$O_{\varphi_{12}} 
\simeq O_{\phi'} \simeq 1/\sqrt{2}$.

\vspace{0.3cm}
$\bullet$ When {\it all} the angles ($s_{13}$, 
$s^L_{12}$ and $s^L_{13}$) are much 
smaller than $\sqrt{m_{\nu_1} m^2_{\nu_2}/m^3_{\nu_3}}$,
the term proportional to $m_{\nu_1}/m_{\nu_2}$ dominates 
in eq.(\ref{theta12-13_small}) and the CP asymmetry reads
\bea
\epsilon &\simeq& \frac{3 y_1^2}{4 \pi} 
\left(\frac{m_{\nu_1}}{m_{\nu_2}}\right) 
\sin(\phi-\phi') ~~.
\eea
In this limit, the results are identical as in the corresponding
limit in the case $V_L=1$,
and there is a single ``leptogenesis phase'', $\phi -\phi'$. So, 
the normalized contributions to the CP asymmetry 
from $\phi$ and $\phi'$ are equal to $1/\sqrt{2}$, while the contributions
from the rest of the phases vanish. The numerical analysis
yield a plot that is very similar to Fig.1, lower right, where
the role of $O_{\delta}$ is played by either $O_{\varphi_{12}}$, 
$O_{\varphi_{13}}$ or $O_{\delta}$.

\vspace{0.3cm}
$\bullet$ Lastly, in the situations where 
the two terms in eq.(\ref{theta12-13_small})
are comparable, the analysis is very involved, since in principle 
there are four independent ``leptogenesis phases'', namely, 
$2 \delta -\phi'$, $2 \varphi_{12} -\phi'$, 
$2 \varphi_{13} -\phi'$ and $\phi-\phi'$. So, the CP asymmetry
receives contributions from the five phases, and in general they
are comparable in size. Hence, it is very difficult to extract
any general conclusion for this case.

\subsection{$|W_{12}|^2m_{\nu_2}/m_{\nu_3} < |W_{13}|^2$}
\label{7.2}

For simplicity, and since the number of phases and angles 
involved is rather large, we will set one of the angles in 
$V_L$ equal to zero, say $s^L_{13}=0$, so the phase $\varphi_{13}$
becomes irrelevant. Since $s^L_{12}$ and $s^L_{13}$ appear in a
similar way in the formulas, one can qualitatively derive
the result when $s^L_{13}$ is different from zero. 
With this choice, we are left with
only two angles, the CHOOZ angle, $s_{13}$, and one angle in $V_L$,
$s^L_{12}$. 
The limit we are studying in this section requires $s^L_{12}$ larger
than $\sim 0.1$. Then, using the experimental bound on the CHOOZ angle
and that our phases are generically of order 1, 
the denominator can be expanded as
\bea
\label{invW13}
\frac{1}{|W_{13}|^2} \simeq \frac{2}{(s^L_{12})^2} 
\left(1+2 \sqrt{2} \frac{c^L_{12}}{s^L_{12}} s_{13} 
\cos(\delta+\varphi_{12}) \right) ~~.
\eea

Hence, the CP asymmetry can be approximated by
\bea
\epsilon &\simeq& -\frac{3 y_1^2}{4 \pi}
\frac{(c^L_{12})^3}{(s^L_{12})^5} \left\{
\left(\frac{s^L_{12}}{c^L_{12}}\right) 
\left[ -2  \sin(2\varphi_{12}+\phi')
+2 \sqrt{2} \left(\frac{s^L_{12}}{c^L_{12}}\right) \sin(\varphi_{12}+\phi')
-\left(\frac{s^L_{12}}{c^L_{12}}\right)^2\sin \phi' \right]
\right. \nonumber \\
& &\left. +s_{13} 
\left[
2\sqrt{2} [
\sin(\delta-\varphi_{12}-\phi')-3\sin(\delta+3\varphi_{12}+\phi')]
\right. \right.\\ \nonumber
& &\left. \left. -4 \left(\frac{s^L_{12}}{c^L_{12}}\right) [\sin(\delta-\phi')
-3 \sin(\delta+2 \varphi_{12}+\phi')]
 +\sqrt{2} \left(\frac{s^L_{12}}{c^L_{12}}\right)^2 
\left[2 \sin(\delta-\varphi_{12}-\phi')
\right. \right. \right. \\ \nonumber
& &\left. \left. \left.
+\sin(\delta+\varphi_{12}-\phi')
-3\sin(\delta+\varphi_{12}+\phi')\right]
-2 \left(\frac{s^L_{12}}{c^L_{12}}\right)^3  \sin(\delta -\phi')
\right]\right\} ~~.
\eea
This expression is rather cumbersome and it is difficult to 
extract information from it. It is not possible in general
to identify the ``leptogenesis phase'', although it is clear
that leptogenesis depends mainly on $\phi'$ and $\varphi_{12}$,
whereas the dependence on $\delta$ is weaker.

In Fig.3 we show the numerical results for this case. As usual, 
we show a triangle with the meaning of the distances
to the different sides (upper left plot), 
and density plots of the ``phase overlaps'' for
the mass hierarchy $m_{\nu_1}:m_{\nu_2}:m_{\nu_3} = 10^{-2}:0.1:1$,
taking random values
for the phases between 0 and $2 \pi$. In the upper right (lower left)
plot we show the results for $\tan \theta^L_{12}=0.5~(1)$ and $s_{13}=0.1$.
In both plots, the points are concentrated close to the base of
the triangle (that corresponds to $O_{\delta}$ small), due to
the small value of the CHOOZ angle. In the plot corresponding
to $\tan \theta^L_{12}=0.5$ the points are concentrated around
the center of the base, whereas for $\tan \theta^L_{12}=1$, they
are spread all over the base. This can be understood from the dependence
of $\epsilon$ on $\cot \theta^L_{12}$. For $\tan \theta^L_{12}=0.5$,
the terms with both $\varphi_{12}$ and $\phi'$ are the dominant ones,
so $C_{\phi' \varphi_{12}}$ is the largest contribution to the CP asymmetry.
On the other hand, when $\tan \theta^L_{12}=1$ these terms
are comparable to the one with $\sin \phi'$, so $\epsilon$ is
dominated by  $C_{\phi' \varphi_{12}}$ and $C_{\phi'}$. Depending 
on the value of $\phi'$ the points spread along the basis of
the triangle. In the lower right plot we show the numerical 
results for  $\tan \theta^L_{12}=0.5$ and $s_{13}=0.01$. The plot
is similar to the one with  $s_{13}=0.1$ but with an even smaller
value of $O_{\delta}$.
\begin{figure}
\centerline{\hbox{
\psfrag{Od}{\Large $O^2_{\delta}$}
\psfrag{Op12}{\Large $O^2_{\varphi_{12}}$}
\psfrag{Opp}{\Large $O^2_{\phi'}$}
\psfig{figure=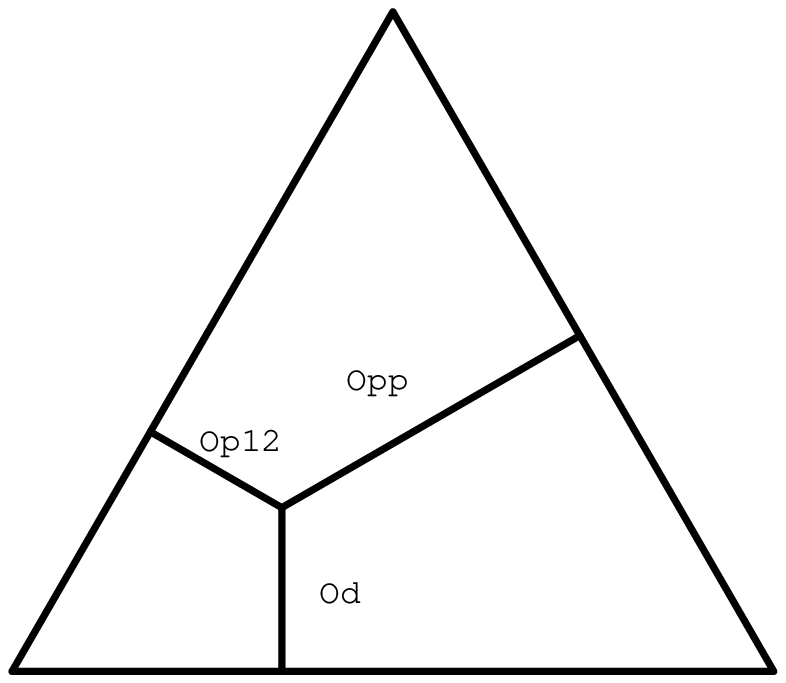,height=10.4cm,width=11.4cm,bbllx=1.cm,%
bblly=0.1cm,bburx=12.cm,bbury=9.1cm}
\psfig{figure=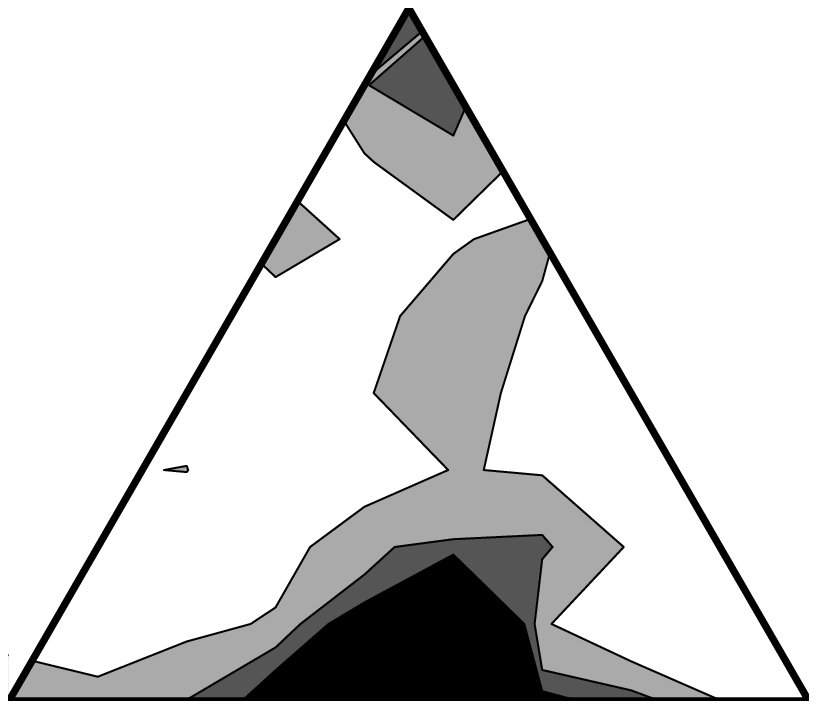,height=10cm,width=11cm,bbllx=3.cm,%
bblly=0.cm,bburx=14.cm,bbury=9.cm}}}
\centerline{\hbox{
\psfig{figure=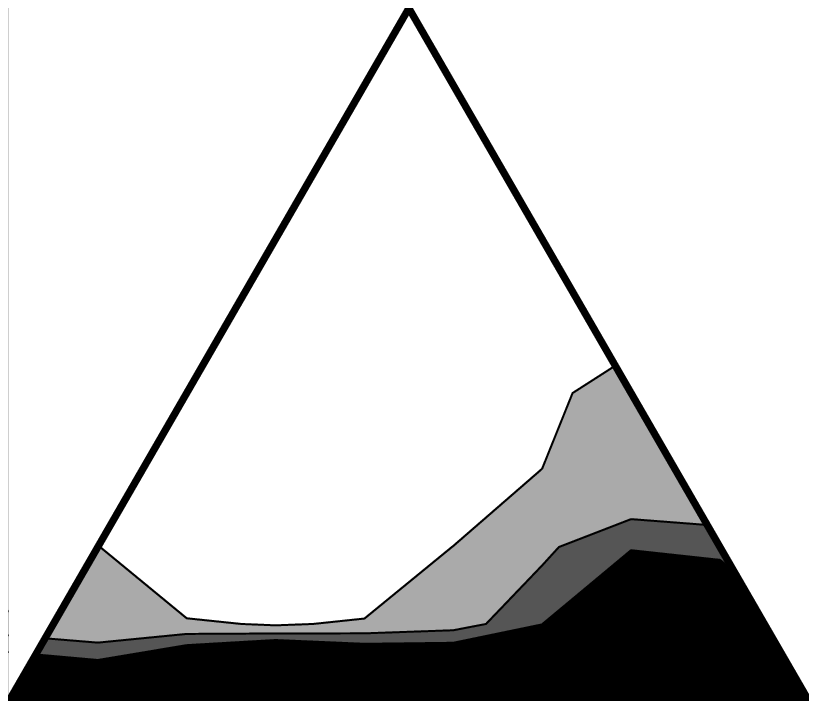,height=10cm,width=11cm,bbllx=1.cm,%
bblly=0.cm,bburx=12.cm,bbury=9.cm}
\psfig{figure=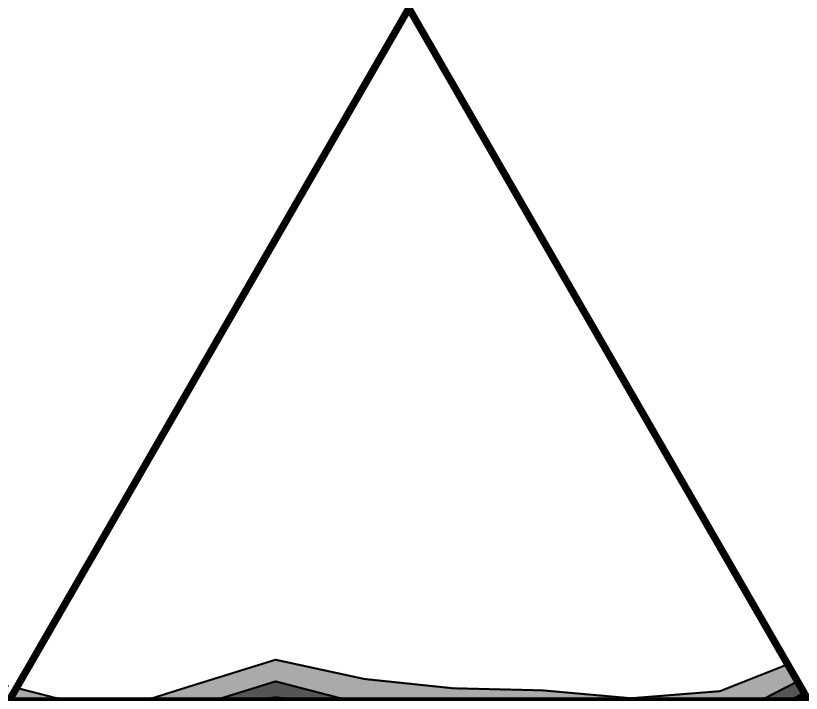,height=10cm,width=11cm,bbllx=3.cm,%
bblly=0.cm,bburx=14.cm,bbury=9.cm}}}
\caption
{\footnotesize 
The same as fig.1 for different situations where 
$|W_{12}|^2m_{\nu_2}/m_{\nu_3} < |W_{13}|^2$. The upper left
plot indicates the meaning of the distances to the different
sides of the triangle. The rest of the triangles show
density plots of the ``phase overlaps'' for random values
of the phases, $s^L_{13}=0$ and $\tan \theta^L_{12}=0.5$, $s_{13}=0.1$ 
(upper right), $\tan \theta^L_{12}=1$, $s_{13}=0.1$ (lower left), and
$\tan \theta^L_{12}=0.5$, $s_{13}=0.01$ (lower right). The darkest regions
correspond to the largest density of points. (See sect. 7.2 for details.)
}
\label{fig3}
\end{figure}

\section{Summary and Discussion}
\label{9}

If neutrino masses are due to the seesaw mechanism, then the
heavy right-handed neutrinos can generate a lepton 
asymmetry in the early Universe when 
they decay out of equilibrium, if they have CP
violating couplings.  Such complex couplings in
the high energy  parameters could induce three phases in the light
neutrino sector, called $\phi$, $\phi'$ and $\delta$
(these are the phases that appear in the MNS matrix; see eqs. (\ref{UV})
and (\ref{Vdef})).
 Upcoming experiments may
be sensitive to two of these phases: the Dirac phase $\delta$
could be measured at a neutrino factory, whereas the Majorana phase
$\phi'$ might have some observable effects 
in neutrinoless double beta decay. 
In this paper, we are interested in 
the relative importance of
the phases $\phi'$ and $\delta$ for leptogenesis. 

To address this issue, we use a parametrization of 
the seesaw in terms of weak scale variables:
the light neutrino masses, the MNS matrix, the eigenvalues of the
neutrino Yukawa matrix, and a unitary matrix $V_L$. We assume a hierarchical
light neutrino spectrum, with the lightest neutrino mass
of order $m_{\nu_3}/100$, and an MNS matrix that corresponds
to the LAMSW solution to the solar neutrino problem. 
The matrix $V_L$ is related
to the off-diagonal (lepton flavour violating) elements of the slepton
mass matrix, and contains three
phases, two of which ($\varphi_{12}, \varphi_{13}$) are relevant
for our calculation. 
It is important to use a parametrization in terms
of ``physical'' weak scale phases;
this  is discussed in section \ref{5}.

In the parameter space we are interested in, we find
a simple analytic approximation for  the lepton asymmetry $\epsilon_1$, 
 produced in the decay of the {\it lightest}
right-handed neutrino $\nu_{R_1}$:
\beq
\epsilon_1 \simeq 
\frac{3 y_1^2}{8 \pi \sum_j |W_{1j}|^2  m_{\nu_j}^2} 
 {\rm Im} \left\{ \frac{ \sum_k W_{1k}^{2} m_{\nu_k}^3 }
{ \sum_n W_{1n}^{2} m_{\nu_n} } \right\} 
\label{epsapprox2}
\eeq
(see eq. (\ref{epsapprox})). In this equation, $y_1$
is  the smallest eigenvalue of the neutrino
Yukawa matrix, and $W$ is the unitary transformation
from the basis where the $\nu_L$ mass matrix is diagonal to the basis
where ${\bf Y^{\dagger}_{\nu}} {\bf Y_\nu}$  is diagonal: $W_{1n} = 
[V_L]_{1m}  [U]_{mn}$ where $V_L$ and
$U$ are defined in eqs.  (\ref{UV}), (\ref{Vdef})
and (\ref{VLdef}). 

We are interested in the {\it relative} importance of the phases 
$\phi'$ and $\delta$ for $\epsilon$. That is, we do not
discuss whether we get $\epsilon$ large enough,
which is essentially controlled by real parameters,
such as $y_1$. We assume that the observed
baryon asymmetry  is produced in the out of equilibrium decay of
$\nu_{R_1}$ and study how important could be $\delta$ or
$\phi'$ for the CP asymmetry. In other words,
if we suppose that $\epsilon$ is of
the correct size, what fraction 
of $\epsilon$ is due to $\delta$
or $\phi'$?

We Fourier expand $\epsilon$ on the five relevant phases
of our low-energy parametrization
\bea
\epsilon   =   \sum_{j,k,l,m,n} A_{j k l m n} \sin(j \delta 
+ k \phi  + l \phi' + m \varphi_{12} + n \varphi_{13}) ~~,
\eea
and then divide the sum into five components, one
due to each phase.  In $\epsilon_{\delta}$, which is 
 the component due to $\delta$,
we put all the terms from the Fourier expansion that
are $\propto \sin (j \delta)$. We define $C_{\delta \alpha} $ 
to be the sum of all  the terms 
$\propto \sin (j \delta + n \alpha)$, and divide it
equally between $\epsilon_\delta$ and $\epsilon_\alpha$ --- 
that is, we add  $\frac{1}{2} \sum_\alpha C_{\delta \alpha}$ to
$\epsilon_\delta$. We also
add the terms $\propto \sin( j \delta + n \alpha + m \beta)$, multiplied
by 1/3, and so on $(\alpha$ and $\beta$ are one of $\{ \phi, \phi',
\varphi_{12}, \varphi_{13} \}$). 
This procedure is described in more detail in section \ref{5}, and 
the formula for $\epsilon_{\delta}$ can be found
in eq.(\ref{epsdelta}).
Then we define a normalized ``fraction of $\epsilon$ due
to $\delta$'', which we call the overlap between
the leptogenesis phase and $\delta$, as
\beq
O_{\delta} = \frac{|\epsilon_{\delta}|}
{\sqrt{ \sum_\alpha \epsilon_\alpha^2}} ~~.
\eeq
The magnitude of $O_\delta$ or $O_{\phi'}$ depends on
five phases and three unknown real parameters: the
CHOOZ angle, $\theta_{13}$, and two angles from $V_L$, that
is related to radiative decays. In the numerical calculation,
we fix the hierarchy of light neutrino masses, the
angles and assign random values (linearly
distributed) to the phases between $0$ and $2 \pi$. The
numerical results are shown in density plots in the  
$O_\delta - O_{\phi'}$ space.

We present results for two representative
cases.
In section \ref{6} we discuss the $V_L = 1$ case, where the three
relevant phases
are $\delta$, $\phi'$ and $\phi$ (the phases of the
MNS matrix), and in Section \ref{7} we allow for two non-zero
angles in $V_L$. For the sake of clarity in the presentation, 
in Section \ref{7} we analyze simplified scenarios
to reduce the number of phases involved to three.
Since the definition of overlap satisfies the
identity  $\sum O_\alpha^2 = 1$, a convenient
way of showing the results is by using a triangular
plot, where the distance to each side of the triangle
corresponds to $O^2_\alpha$.

The $V_L = 1$ model should be a good approximation 
when the angles in $V_L$ 
are smaller than the CHOOZ angle. It can be checked
from eqs.(\ref{W(49)})
and (\ref{epsapprox2}) that the Dirac phases
$\delta$, $\varphi_{12}$ and $\varphi_{13}$ appear
in $\epsilon$ multiplied by the sine of a real
angle ($\theta_{13}$, $\theta_{12}^L$
and $\theta_{13}^L$, respectively). If $\theta_{12}^L$ 
and $\theta_{13}^L $ are much smaller than the CHOOZ angle $\theta_{13}$, then
$\varphi_{12}$ and $\varphi_{13}$ are less important
for $\epsilon$ than $\delta$. 

In Section \ref{7} we analyze the situation where there
are angles in $V_L$  larger than
the CHOOZ angle. The angles of $V_L$  are related
to the branching ratios for $\ell_j \rightarrow \ell_i \gamma$,
as discussed after eq. (\ref{softafterRG}) : $
BR(\ell_j \rightarrow \ell_i \gamma) \propto y_k^2 | [V_L]_{kj}^*
[V_L]_{ki}|^2$. 
Current limits on  $\tau \rightarrow \mu \gamma$,
 $\tau \rightarrow e \gamma$, and  $\mu \rightarrow e \gamma$
are  satisfied if all
angles $\theta^L_{ij} \lsim .1$. However, present bounds
and anticipated improvements on all three branching ratios 
can be satisfied  if {\it e.g.}  $\theta^L_{13}, \theta^L_{23} \simeq  0$
and  $\theta^L_{12} \sim 1$. So it is phenomenologically possible to 
have at least one large angle.  The associated phase could then be
important for leptogenesis; this possibility is
studied in that section.  

The approximate expression for $\epsilon$ given in eq.(\ref{epsapprox2}),
shows that for generic $W_{1j}$, $\epsilon \propto$ 
${\rm Im} (W_{12} W^*_{13})^2 $.
That is, the terms proportional to  $m_{\nu_1}$
can be neglected when $W_{12} > W_{11} \sqrt{m_{\nu_1}/m_{\nu_2}}$
and  $W_{13} > W_{11} \sqrt{m_{\nu_1} m_{\nu_2}^2 /m_{\nu_3}^3}$,
as discussed in section \ref{7}.
Recall that  $W_{13}$ contains terms $\sim \sin \theta_{13}$, $
 \sin \theta^L_{12}$, $\sin \theta^L_{13}$. We set 
 $\sin \theta^L_{13}$ to zero, to ensure that  
$\ell_j \rightarrow \ell_i \gamma$ constraints are satisfied,
and because the functional dependence of $\epsilon$ on 
$ \sin \theta^L_{12}$ and  $\sin \theta^L_{13}$ is similar
\footnote{ this is because $\nu_3 \simeq (\nu_\tau + \nu_\mu)/\sqrt{2}$}.
So the case studied in that section has three phases: 
$\delta $, $\phi'$ and $\varphi_{12}$.

The CP asymmetry $\epsilon$ usually  depends on the interference between
at least two phases. Our results can be divided into three 
representative cases, according to the neutrino masses
and low energy mixing angles. The phases relevant to leptogenesis can be
1) the light majorana phases $\phi$ and $\phi'$, or 2) the neutrino
factory phase $\delta$ and $\phi'$, or 3) $\phi'$ and  phase(s)
from the slepton mass matrix. The first two cases were discussed
in section \ref{6}, and the third in section \ref{7}.
We outline here the observational consequences of each case.

The majorana phases $\phi$ and $\phi'$ are the relevant phases
for leptogenesis (equivalently $O_{\phi}$ and $O_{\phi'}$
are large) when
\beq
\theta_{13}, 
 ~ \theta^L_{1j} \ll  \sqrt{\frac{m_{\nu_2}^2 m_{\nu_1}}{m_{\nu_3}^3}}  ~~~~, ~ (O_\phi , O_{\phi'} ~
{\rm large})
\label{case1}
\eeq
where $j = 1,2$. For $m_{\nu_1} \sim .01 m_{\nu_3}$, 
\footnote{ We use $m_{\nu_1} = .01 m_{\nu_3}$ to estimate numerical
upper bounds on the angles. The parameter space
where this scenario obtains shrinks down as $m_{\nu_1}$ decreases.}
these conditions imply a CHOOZ angle $\theta_{13} < .01$,
 $BR(\tau \rightarrow e \gamma) \lsim 10^{-10}$
and  $BR(\mu \rightarrow e \gamma) \lsim 10^{-4} 
BR(\tau \rightarrow \mu \gamma) $. \footnote{In this discussion, we
approximate (!) $BR(\ell_j \rightarrow \ell_i \gamma) \sim
10^{-7} y_k^2 |V_{Lkj}^* V_{Lki}|^2 \left(\frac{\tan \beta}{10} \right)^2
 \left(\frac{300{\rm GeV}}{m_{susy}} \right)^4$, so these estimates are
for flavour only and should be taken with large crystals of
salt.}  
So if this scenario is realised, then
with forseeable sensitivities, 
 $\tau \rightarrow \mu \gamma$ and  $\mu \rightarrow e \gamma$  
could be observed,  but  $\theta_{13}$ and $\tau  \rightarrow  e \gamma$
will not be seen. 
If $m_{\nu_1}$ was measured in $0 \nu 2 \beta$ decay,
then sufficient conditions (eqn \ref{case1}) could be determined:
knowing $m_{\nu_1}$ fixes how small $\theta_{13} $ and
$\theta_{1j}^L$ must be for their associated phases to be
irrelevant. 

Notice that $\tau \rightarrow \mu \gamma$ has little
impact on the importance of $\delta$ for leptogenesis. 
This is because it is mostly related to $\theta_{23}^L$,
the angle from $V_L$ which does not affect $\epsilon$ in our
parametrisation.

A second possibility, where $\delta$ would be important for
leptogenesis, arises if
\beq
\theta^L_{1j} , \sqrt{ \frac{m_{\nu_2}^2 m_{\nu_1}}{m_{\nu_3}^3}} 
  \ll \theta_{13}  ~~~~, ~ (O_\delta , O_{\phi'} ~
{\rm large})
\label{case2}
\eeq
This in the interesting case for the neutrino factory; 
it would arise for $\theta_{13}> .01 \sqrt{\frac{m_{\nu_1}}{.01m_{\nu_3}}}$
and $\theta_{1j}^L < \theta_{13}$. For $m_{\nu_1} \sim .01 m_{\nu_3}$,
it corresponds to a CHOOZ
angle accessible to the neutrino factory, $\tau \rightarrow e \gamma$
below anticipated sensitivities ($BR(\tau \rightarrow e \gamma) <10^{-9}$),
 and $BR(\mu \rightarrow e \gamma) \lsim \theta_{13}^2 
BR(\tau \rightarrow \mu \gamma) $.
 Case 2 (eq \ref{case2}) corresponds
to $O_\delta$ large, and therefore $\tau \rightarrow e \gamma$
unobservably small.

A final possibility is that the phases of the slepton
mass matrix are more important
for leptogenesis than $\delta$. This corresponds to
\beq
\theta_{13}  , \sqrt{ \frac{m_{\nu_2}^2 m_{\nu_1}}{m_{\nu_3}^3}} 
  \ll \theta^L_{1j}  ~~~~, ~ (O_{\varphi_{1j}} , O_{\phi'} ~
{\rm large})
\label{case3}
\eeq
An example of this case would be $ \theta_{13} \sim .01$---so
detectable at the neutrino factory, and 
$ \sin \theta^L_{13} \simeq 1/\sqrt{5}$---which
would imply $BR(\tau \rightarrow e \gamma) \sim 10^{-8}$.
This example is plotted in figure \ref{fig3} (if
$\{ \theta_{13}^L, \varphi_{13} \}$ and 
$\{ \theta_{12}^L, \varphi_{12} \}$ are interchanged). 
If  $\tau \rightarrow e \gamma $ is observed, 
then we are in case 3, $ O_{\varphi_{1j}} > O_\delta$
and the slepton phases are
probably more important for leptogenesis than $\delta$. 
Notice however, that lepton flavour violating
processes could be small even in this case,
where $\delta$ is not important for leptogenesis. 
 The 1-2 angle of $V_L$ can be
large, without inducing observable  
$\ell_j \rightarrow \ell_i \gamma$  rates, because it induces
lepton flavour violating slepton masses proportional 
to $y^2_2$. 
Such simultaneously small
lepton flavour violation and  $O_\delta$ 
could be disfavoured by model-building, 
because it requires $\theta^L_{13} \ll \theta^L_{12}$.

To summarise these three cases:
we find that the neutrino factory phase $\delta$ {\it can} be
important for leptogenesis. When the CHOOZ
angle is large enough to detect CP violation at
neutrino factories (which requires $\theta_{13} \gsim .01$),
and the lepton flavour violating branching
ratios are vanishingly small 
($\theta^L_{12},~ \theta^L_{13}\lsim \theta_{13}$),
 then $\delta$ contributes significantly to the 
 leptogenesis phase ($|O_\delta|^2 \gsim .3 $).
This can be seen from our low-energy approximation
to $\epsilon$, eq. (\ref{epsapprox2}): if
$W_{13}$ is not too small, $\epsilon$ depends 
on the phase difference between $W_{12}$ and $W_{13}$,
and if  $\theta_{13} \gsim  \theta^L_{12},~ \theta^L_{13}$,
the phase of $W_{13}$ is $\delta$.
Although $\delta$  appears always suppressed by
the CHOOZ angle, which is small, it plays
an important role, because $W_{13}$ is multiplied
by the largest neutrino mass
$m_{\nu_3}$.
The phase $\delta$ is unlikely to be important for leptogenesis
if either the CHOOZ angle is small ($ \theta_{13} < .01$
---see fig \ref{fig1}),
or if the $\tau \rightarrow e \gamma$ or
$\mu \rightarrow e \gamma$ branching ratios are large
($  \theta^L_{12},~ \theta^L_{13} >  \theta_{13}$ ---see fig \ref{fig3}).

We find that the Majorana phase $\phi'$ is (almost)
always important for leptogenesis. For instance, the fraction of
$\epsilon$ that is due to $\phi'$,
 $O_{\phi'}$,  is significant in 
all the cases we have studied. 
Algebraically, the reason is that the main contribution to
$\epsilon$ in eq.(\ref{epsapprox2}) is
generically proportional to  $W_{12}$, that is proportional 
to $e^{i \phi'}$, unless some cancellations 
occur.  The contribution from any of
the other phases can be suppressed by sending a small parameter
 to zero. The Majorana phase of $m_{\nu_1}$, $\phi$, becomes
unimportant as $m_{\nu_1} \rightarrow 0$, and the three
Dirac phases $\delta$, $\varphi_{12}$  and $\varphi_{13}$
multiply angles which are positively small ($s_{13}$) or
believed to be small ($s^L_{12}$, $s^L_{13}$). The
contribution of $\phi'$, on the other hand,
is consistently significant.

It is interesting to study how closely related are the
Majorana phases of the light neutrinos to the Majorana
phases of the heavy right-handed neutrinos.
It is well known  that
when neutrinos have Majorana masses, there is 
CP violation even in the two generation model. This suggests that
the Majorana phases of the $\nu_R$ sector
could be  more important for leptogenesis than the
Dirac phase, because they can  contribute to the CP
asymmetry $\epsilon$ suppressed by mixing between
only two of the $\nu_R$, rather than mixing among the 
three $\nu_R$, as required for the Dirac phase.
However, there is no symmetry-based distinction between  Majorana
phases and Dirac phases. The high scale Majorana
phases are functions of all the weak scale
parameters---the real ones as well
as all the Dirac and Majorana phases.
So the reason $\phi'$ is important is not
that  low energy Majorana phases determine the
high scale Majorana phases. 
 One can check
from the formulae in section \ref{4} that
$\phi'$ is usually significant because
it multiples a not-very-small mass, rather than a
(possibly) small mixing angle.

We do not find a simple correlation between the
sign of low energy phases and the sign of the
CP asymmetry $\epsilon$. Such a correlation
would be interesting, and exists in certain models
\cite{Buchmuller:2001dc}. However, in our bottom-up
approach, $\epsilon$ is usually proportional to
phase differences ($\epsilon \sim \sin(j \alpha + n \beta)$),
as can be seen from  eq. (\ref{epsapprox2})
and the various limiting cases discussed in
sections \ref{6} and \ref{7}.

In summary, we have studied the relative importance of
low energy phases for leptogenesis. Using a
parametrization of the seesaw mechanism in terms of weak 
scale variables, we express the CP asymmetry 
produced in the decay of the lightest
$\nu_R$ as a function of the ``neutrino factory phase''
$\delta$, the ``neutrinoless double beta decay phase''
$\phi'$, and three other ``physical'' weak scale
phases.  We introduce a way
of splitting $\epsilon$ into contributions due to the 
different phases, $\delta$, $\phi'$, etc. We assume that
$\epsilon$ is big enough to be responsible for
the observed baryon asymmetry, and compare
the relative size of the different contributions.
We find that $\phi'$ is generically important for
leptogenesis. The importance
of $\delta$ depends on the mixing angles
of the slepton sector. If these are smaller
than the CHOOZ angle, then $\delta$ makes
a significant contribution to leptogenesis.

\subsection*{Acknowledgments}
We thank Felipe Joaquim, 
Michael Pl\"umacher and Graham Ross for discussions, and
 Wilfred Buchm\"uller, F Schrempp and particularily Laura Covi  for
comments and suggestions.

\subsection*{Addendum}
After this work was completed, a related analysis
appeared \cite{Ellis:2002xg}.

\section*{Appendix}

In the appendix we explain the numerical procedure that
we have followed to compute the contributions
to the CP asymmetry from the different phases.
For the sake of clarity we will only present the 
procedure we followed for the case $V_L=1$, where
only the phases $\delta$, $\phi$ and $\phi'$ were
relevant. The extension to the general case is
straight-forward.

Our starting point to compute the contributions was
the Fourier expansion of the CP asymmetry
\bea
\label{fourier_app}
\epsilon   =   \sum_{j,k,l} A_{j k l} \sin(j \delta+k \phi+l \phi')
= C_{\delta}+C_{\phi}+C_{\phi'}+
 C_{\delta \phi} +C_{\delta \phi'} +C_{\phi \phi'} +
C_{\delta \phi \phi'}
\eea
with $C_{\alpha}$, $C_{\alpha \beta}$ and $C_{\delta \phi \phi'}$ as
in eqs.(\ref{Cdelta})-(\ref{Cdeltaphiphip}).
From the periodicity of $\epsilon$ it is apparent that 
\bea
C_{\delta}&=&\frac{1}{(2 \pi)^2} 
\int_{-\pi}^{\pi} d \phi ~ d\phi' ~\epsilon \nonumber \\
C_{\phi}&=&\frac{1}{(2 \pi)^2} 
\int_{-\pi}^{\pi} d \delta ~ d\phi' ~\epsilon \\
C_{\phi'}&=&\frac{1}{(2 \pi)^2} 
\int_{-\pi}^{\pi} d \delta ~ d\phi ~\epsilon \nonumber 
\eea
\bea
C_{\delta \phi}&=&\frac{1}{(2 \pi)} 
\int_{-\pi}^{\pi} d \phi' ~ \epsilon -(C_{\delta}+C_{\phi'}) \nonumber \\
C_{\delta \phi'}&=&\frac{1}{(2 \pi)} 
\int_{-\pi}^{\pi} d \phi ~ \epsilon -(C_{\delta}+C_{\phi})  \\
C_{\phi \phi'}&=&\frac{1}{(2 \pi)} 
\int_{-\pi}^{\pi} d \delta ~ \epsilon -(C_{\phi}+C_{\phi'}) \nonumber 
\eea
\bea
C_{\delta \phi \phi'}&=& \epsilon-(C_{\delta}+C_{\phi}+C_{\phi'}+
C_{\delta \phi}+C_{\delta \phi'}+C_{\phi \phi'}) ~~.
\eea

These integrals can be computed numerically, thus giving 
the different contributions to the CP asymmetry.  This
avoids difficulties with the points where the approximation
eq. (\ref{epsapprox}) breaks down. However, it is
also interesting to solve this integrals analytically,
to cross-check the results we  obtained in Section 6.
The results for the double integrals are
\bea
C_{\delta} \simeq 0 ~~~~~ 
C_{\phi}\simeq 0 ~~~~~
C_{\phi'} \simeq 0.
\eea
On the other hand, the results for the single integrals is 
more involved and depends
on the particular point of the parameter space.
These integrals can be computed using the residue theorem;
the number of poles inside the unit circle depends on the values of the
phases and other neutrino parameters, especially on the CHOOZ angle, 
hence the dependence of the result on the chosen parameters.
However, some care must be exercised in using the
residue theorem, because there can be poles in 
 eq. (\ref{epsapprox}) at points where
the  approximation breaks down. Such poles
must be neglected. 

The results
for the single integrals
are different depending on the CHOOZ angle. We 
consider three possibilities:

$\bullet$ When the CHOOZ angle is close to the
experimental upper limit, or to be precise, when
\bea
&|m_{\nu_1}~ c_{13}^2 c_{12}^2 ~e^{i \phi}+
m_{\nu_3} ~s_{13}^2 ~ e^{2 i \delta}|< m_{\nu_2} ~c_{13}^2 s_{12}^2 & 
\nonumber \\
&|m_{\nu_2} ~c_{13}^2 s_{12}^2~ e^{ i \phi'}+
m_{\nu_3} ~s_{13}^2 ~ e^{2 i \delta}|>m_{\nu_1}~ c_{13}^2 c_{12}^2 & \\
&|m_{\nu_1}~ c_{13}^2 c_{12}^2 ~e^{ i \phi}+
m_{\nu_2} ~c_{13}^2 s_{12}^2~ e^{ i \phi'}|<
m_{\nu_3} ~s_{13}^2 ~~, &\nonumber
\eea
the single integrals read
\bea
C_{\delta \phi} &\simeq& 0 \nonumber \\
C_{\delta \phi'} &\simeq& -\frac{3 y_1^2}{8 \pi D} ~
{\rm Im} \left\{\frac{
m_{\nu_2}^3 ~c_{13}^2 s_{12}^2~ e^{ i \phi'}+
m_{\nu_3}^3 ~s_{13}^2 ~ e^{2 i \delta}}
{m_{\nu_2} ~c_{13}^2 s_{12}^2~ e^{ i \phi'}+
m_{\nu_3} ~s_{13}^2 ~ e^{2 i \delta}} \right\} 
\simeq  -\frac{3 y_1^2}{2 \pi} ~
\left( \frac{m_{\nu_3}}{m_{\nu_2}}\right)^3 s_{13}^2 
\sin(2 \delta -  \phi')  \nonumber \\
C_{\phi \phi'} &\simeq& 0  ~~,
\eea
where $D$ was defined after eq.(\ref{eps_analytic1}). This result,
coincide with eq.(\ref{CHOOZ_large}), that was obtained
using a completely different method.

$\bullet$ When the CHOOZ angle is very small, or when the conditions
\bea
&|m_{\nu_1}~ c_{13}^2 c_{12}^2 ~e^{ i \phi}+
m_{\nu_3} ~s_{13}^2 ~ e^{2 i \delta}|< m_{\nu_2} ~c_{13}^2 s_{12}^2 & 
\nonumber \\
&|m_{\nu_2} ~c_{13}^2 s_{12}^2~ e^{ i \phi'}+
m_{\nu_3} ~s_{13}^2 ~ e^{2 i \delta}|<m_{\nu_1}~ c_{13}^2 c_{12}^2 & \\
&|m_{\nu_1}~ c_{13}^2 c_{12}^2 ~e^{ i \phi}+
m_{\nu_2} ~c_{13}^2 s_{12}^2~ e^{ i \phi'}|>
m_{\nu_3} ~s_{13}^2 &\nonumber
\eea
are fulfilled, the results for the single integrals are
\bea
C_{\delta \phi} &\simeq& 0 \nonumber \\
C_{\delta \phi'} &\simeq& 0 \\
C_{\phi \phi'} &\simeq& -\frac{3 y_1^2}{8 \pi D} ~
{\rm Im} \left\{\frac{
m_{\nu_1}^3 ~c_{13}^2 c_{12}^2 ~e^{ i \phi}+
m_{\nu_2}^3 ~c_{13}^2 s_{12}^2~ e^{ i \phi'}}
{m_{\nu_1}~ c_{13}^2 c_{12}^2 ~e^{ i \phi}+
m_{\nu_2} ~c_{13}^2 s_{12}^2~ e^{ i \phi'}} \right\} 
\simeq  \frac{3 y_1^2}{4 \pi} ~\frac{m_{\nu_1}}{m_{\nu_2}}
\sin (\phi-\phi') ~~. \nonumber
\eea
This result is identical to the result obtained using 
series expansions in Section 6, eq.(\ref{CHOOZ_small}).

$\bullet$ For intermediate values of the CHOOZ angle, it 
is usually the case that 
\bea
&|m_{\nu_1}~ c_{13}^2 c_{12}^2 ~e^{ i \phi}+
m_{\nu_3} ~s_{13}^2 ~ e^{2 i \delta}|< m_{\nu_2} ~c_{13}^2 s_{12}^2 & 
\nonumber \\
&|m_{\nu_2} ~c_{13}^2 s_{12}^2~ e^{ i \phi'}+
m_{\nu_3} ~s_{13}^2 ~ e^{2 i \delta}|>m_{\nu_1}~ c_{13}^2 c_{12}^2 & \\
&|m_{\nu_1}~ c_{13}^2 c_{12}^2 ~e^{ i \phi}+
m_{\nu_2} ~c_{13}^2 s_{12}^2~ e^{ i \phi}|>
m_{\nu_3} ~s_{13}^2 ~~, &\nonumber
\eea
so the single integrals are
\bea
C_{\delta \phi} &\simeq& 0 \nonumber \\
C_{\delta \phi'} &\simeq& -\frac{3 y_1^2}{8 \pi D} ~
{\rm Im} \left\{\frac{
m_{\nu_2}^3 ~c_{13}^2 s_{12}^2~ e^{ i \phi'}+
m_{\nu_3}^3 ~s_{13}^2 ~ e^{2 i \delta}}
{m_{\nu_2} ~c_{13}^2 s_{12}^2~ e^{ i \phi'}+
m_{\nu_3} ~s_{13}^2 ~ e^{2 i \delta}} \right\}
\simeq  -\frac{3 y_1^2}{2 \pi } ~ 
\left(\frac{m_{\nu_3}}{m_{\nu_2}}\right)^3 s_{13}^2 
\sin(2\delta -  \phi')  \nonumber \\
C_{\phi \phi'} &\simeq&  -\frac{3 y_1^2}{8 \pi D} ~
{\rm Im} \left\{\frac{
m_{\nu_1}^3 ~c_{13}^2 c_{12}^2 ~e^{ i \phi}+
m_{\nu_2}^3 ~c_{13}^2 s_{12}^2~ e^{ i \phi'}}
{m_{\nu_1}~ c_{13}^2 c_{12}^2 ~e^{ i \phi}+
m_{\nu_2} ~c_{13}^2 s_{12}^2~ e^{ i \phi'}} \right\} 
\simeq  \frac{3 y_1^2}{4 \pi } ~\frac{m_{\nu_1}}{m_{\nu_2}} \sin (\phi-\phi')  ~~, 
\eea
that are identical to eq.(\ref{epsdelta2}).


\begin{thebibliography}{222222}
%



\bibitem{Cleveland:1998nv}
R.~J.~Davis, D.~S.~Harmer and K.~C.~Hoffman,
Phys.\ Rev.\ Lett.\  {\bf 20} (1968) 1205.


\bibitem{SK}
Y.~Fukuda {\it et al.}  [Super-Kamiokande Collaboration],
Phys.\ Rev.\ Lett.\ {\bf 81} (1998) 1562,
Phys.\ Rev.\ Lett.\ {\bf 82} (1999) 1810,
Phys.\ Rev.\ Lett.\ {\bf 82} (1999) 2430.


\bibitem{Fukugita:1986hr}
M.~Fukugita and T.~Yanagida,
Phys.\ Lett.\ B {\bf 174} (1986) 45.




\bibitem{Buchmuller:2000wq}
{\it see e.g.} W.~Buchmuller and S.~Fredenhagen,
hep-ph/0001098;
A.~Riotto and M.~Trodden,
Ann.\ Rev.\ Nucl.\ Part.\ Sci.\  {\bf 49} (1999) 35;
V.~A.~Rubakov and M.~E.~Shaposhnikov,
Usp.\ Fiz.\ Nauk {\bf 166} (1996) 493
[Phys.\ Usp.\  {\bf 39} (1996) 461].




\bibitem{Sakharov:1967dj}
A.~D.~Sakharov,
Pisma Zh.\ Eksp.\ Teor.\ Fiz.\  {\bf 5} (1967) 32
[JETP Lett.\  {\bf 5} (1967) 24].

\bibitem{seesaw} 
M. Gell-Mann, P. Ramond and
R. Slansky,  {\em Proceedings of the Supergravity Stony Brook Workshop}, New
York 1979,  eds. P. Van Nieuwenhuizen and D. Freedman; T. Yanagida,  {\em
Proceedinds of the Workshop on Unified Theories and Baryon Number in the
Universe},  Tsukuba, Japan 1979, ed.s A. Sawada and A. Sugamoto;
R. N. Mohapatra, G. Senjanovic,
{\it Phys.Rev.Lett.} {\bf 44} (1980)912, {\it ibid.}
{\it Phys.Rev.} {\bf D23} (1981) 165.
%


\bibitem{Branco:2001pq}
G.~C.~Branco, T.~Morozumi, B.~M.~Nobre and M.~N.~Rebelo,
arXiv:hep-ph/0107164.


\bibitem{Branco:2002kt}
G.~C.~Branco, R.~Gonzalez Felipe, F.~R.~Joaquim and M.~N.~Rebelo,
arXiv:hep-ph/0202030.


\bibitem{Ellis:2001xt}
J.~R.~Ellis, J.~Hisano, S.~Lola and M.~Raidal,
arXiv:hep-ph/0109125.


\bibitem{LR}
A.~S.~Joshipura, E.~A.~Paschos and W.~Rodejohann,
JHEP {\bf 0108} (2001) 029
[arXiv:hep-ph/0105175].
%
W.~Rodejohann and K.~R.~Balaji,
Phys.\ Rev.\ D {\bf 65} (2002) 093009
[arXiv:hep-ph/0201052].


\bibitem{GUT}
M.~S.~Berger and K.~Siyeon,
Phys.\ Rev.\ D {\bf 65} (2002) 053019
[arXiv:hep-ph/0110001].
%
F.~Buccella, D.~Falcone and F.~Tramontano,
Phys.\ Lett.\ B {\bf 524} (2002) 241
[arXiv:hep-ph/0108172].
%
D.~Falcone and F.~Tramontano,
Phys.\ Lett.\ B {\bf 506} (2001) 1
[arXiv:hep-ph/0101151].
%
D.~Falcone and F.~Tramontano,
Phys.\ Rev.\ D {\bf 63} (2001) 073007
[arXiv:hep-ph/0011053].

\bibitem{King:2002nf}
S.~F.~King,
arXiv:hep-ph/0204360.


\bibitem{Davidson:2001zk}
S.~Davidson and A.~Ibarra,
JHEP {\bf 0109} (2001) 013
[arXiv:hep-ph/0104076].


\bibitem{SNOCC}
Q.~R.~Ahmad {\it et al.}  [SNO Collaboration],
Phys.\ Rev.\ Lett.\  {\bf 87} (2001) 071301;
%

\bibitem{Ahmad:2002jz}
Q.~R.~Ahmad {\it et al.}  [SNO Collaboration],
arXiv:nucl-ex/0204008.
%
Q.~R.~Ahmad {\it et al.}  [SNO Collaboration],
arXiv:nucl-ex/0204009.

\bibitem{Bahcall:2002hv}
V.~Barger, D.~Marfatia, K.~Whisnant and B.~P.~Wood,
Phys.\ Lett.\ B {\bf 537} (2002) 179
[arXiv:hep-ph/0204253].
%
A.~Bandyopadhyay, S.~Choubey, S.~Goswami and D.~P.~Roy,
arXiv:hep-ph/0204286.
%
J.~N.~Bahcall, M.~C.~Gonzalez-Garcia and C.~Pena-Garay,
arXiv:hep-ph/0204314.
%
P.~C.~de Holanda and A.~Y.~Smirnov,
arXiv:hep-ph/0205241.
A.~Strumia, C.~Cattadori, N.~Ferrari and F.~Vissani,
arXiv:hep-ph/0205261.
%
G.~L.~Fogli, E.~Lisi, A.~Marrone, D.~Montanino and A.~Palazzo,
arXiv:hep-ph/0206162.


\bibitem{experiments}
B.~T.~Cleveland {\it et al.},
Astrophys.\ J.\  {\bf 496} (1998) 505;
%
Y.~Fukuda {\it et al.}  [Kamiokande Collaboration],
Phys.\ Rev.\ Lett.\ {\bf 77} (1996) 1683;
%
S.~Hatakeyama {\it et al.}  [Kamiokande Collaboration],
Phys.\ Rev.\ Lett.\ {\bf 81} (1998) 2016;
%
W.~W.~Allison {\it et al.},
Phys.\ Lett.\ B {\bf 391} (1997) 491;
%
W.~W.~Allison {\it et al.}  [Soudan-2 Collaboration],
Phys.\ Lett.\ B {\bf 449} (1999) 137;
%
W.~Hampel {\it et al.}  [GALLEX Collaboration],
Phys.\ Lett.\ B {\bf 388} (1996) 384;
%
D.~N.~Abdurashitov {\it et al.},
Phys.\ Rev.\ Lett.\ {\bf 77} (1996) 4708.

\bibitem{Apollonio:1999ae}
M.~Apollonio {\it et al.}  [CHOOZ Collaboration],
Phys.\ Lett.\ B {\bf 466} (1999) 415.

\bibitem{0nbb} see {\it e.g.}
A.Alessandrello et. al, {\it Phys. Lett.} {\bf B486} (2000) 13;
Heidelberg-Moscow Collaboration, {\it Phys. Rev. Lett.} 83 (1999) 41-44;
Heidelberg-Moscow Collaboration, {\it Nucl. Phys.} {\bf A}694  (2001) 
269-294. 

\bibitem{Lobashev:tp}
V.~M.~Lobashev {\it et al.},
Phys.\ Lett.\ B {\bf 460} (1999) 227.
%
C.~Weinheimer {\it et al.},
Phys.\ Lett.\ B {\bf 460} (1999) 219.


\bibitem{nufact}
{ \it see, e.g.}
C. Albright {\it et al.,} hep-ex/0008064 \\
{\it or, on the web: }
http://www.cap.bnl.gov/mumu/mu\_home\_page.html, or
http://muonstoragerings.web.cern.ch/muonstoragerings/


\bibitem{Cervera:2000kp}
A.~Cervera, A.~Donini, M.~B.~Gavela, J.~J.~Gomez Cadenas, P.~Hernandez, O.~Mena and S.~Rigolin,
Nucl.\ Phys.\ B {\bf 579} (2000) 17
[Erratum-ibid.\ B {\bf 593} (2001) 731]
[arXiv:hep-ph/0002108].
%
M.~Freund, P.~Huber and M.~Lindner,
Nucl.\ Phys.\ B {\bf 615} (2001) 331
[arXiv:hep-ph/0105071].
\bibitem{Romanino:1999zq}
A.~Romanino,
Nucl.\ Phys.\ B {\bf 574} (2000) 675
[arXiv:hep-ph/9909425].
%
 J.~Burguet-Castell, M.~B.~Gavela, J.~J.~Gomez-Cadenas, P.~Hernandez and O.~Mena,
Nucl.\ Phys.\ B {\bf 608} (2001) 301
[arXiv:hep-ph/0103258].


\bibitem{Barger:2002vy}
V.~Barger, S.~L.~Glashow, P.~Langacker and D.~Marfatia,
arXiv:hep-ph/0205290.



\bibitem{Elliott:2002xe}
{\it see e.g.} 
S.~R.~Elliott and P.~Vogel,
arXiv:hep-ph/0202264.

\bibitem{Borzumati:1986qx}
F.~Borzumati and A.~Masiero,
Phys.\ Rev.\ Lett.\  {\bf 57} (1986) 961.


\bibitem{Casas:2001sr}
J.~A.~Casas and A.~Ibarra,
Nucl.\ Phys.\ B {\bf 618} (2001) 171
[arXiv:hep-ph/0103065].


\bibitem{Buchmuller:1999cu}
{\it see e.g.}W.~Buchmuller and M.~Plumacher,
Phys.\ Rept.\  {\bf 320} (1999) 329, {\it and references therein}.



\bibitem{Kuzmin:1985mm}
V.~A.~Kuzmin, V.~A.~Rubakov and M.~E.~Shaposhnikov,
Phys.\ Lett.\ B {\bf 155} (1985) 36.



\bibitem{topdown}
J.~Ellis, M.~E.~Gomez, G.~K.~Leontaris, S.~Lola and D.~V.~Nanopoulos,
Eur.\ Phys.\ J.\ C {\bf 14} (2000) 319;
M.~E.~Gomez, G.~K.~Leontaris, S.~Lola and J.~D.~Vergados,
Phys.\ Rev.\ D {\bf 59} (1999) 116009;
J.~L.~Feng, Y.~Nir and Y.~Shadmi,
Phys.\ Rev.\ D {\bf 61} (2000) 113005.
G.~K.~Leontaris and N.~D.~Tracas,
Phys.\ Lett.\ B {\bf 431} (1998) 90;
W.~Buchmuller, D.~Delepine and L.~T.~Handoko,
Nucl.\ Phys.\ B {\bf 576} (2000) 445.
W.~Buchmuller, D.~Delepine and F.~Vissani,
Phys.\ Lett.\ B {\bf 459} (1999) 171;
D.~F.~Carvalho, M.~E.~Gomez and S.~Khalil,
hep-ph/0101250;
J.~Sato and K.~Tobe,
hep-ph/0012333;
S.~F.~King and M.~Oliveira,
Phys.\ Rev.\ D {\bf 60} (1999) 035003;
R.~Barbieri, L.~Hall and A.~Strumia,
Nucl.\ Phys.\ B {\bf 445} (1995) 219.

\bibitem{bottomup}
J.~Hisano, T.~Moroi, K.~Tobe and M.~Yamaguchi,
Phys.\ Rev.\ D {\bf 53} (1996) 2442.
%
J.~Hisano and D.~Nomura,
Phys.\ Rev.\ D {\bf 59} (1999) 116005.
%
J.~Sato, K.~Tobe and T.~Yanagida,
Phys.\ Lett.\ B {\bf 498} (2001) 189.
J.~Hisano, D.~Nomura, Y.~Okada, Y.~Shimizu and M.~Tanaka,
Phys.\ Rev.\ D {\bf 58} (1998) 116010;
J.~Hisano, D.~Nomura and T.~Yanagida,
Phys.\ Lett.\ B {\bf 437} (1998) 351;
J.~Hisano, T.~Moroi, K.~Tobe and M.~Yamaguchi,
Phys.\ Lett.\ B {\bf 391} (1997) 341;
P.~Ciafaloni, A.~Romanino and A.~Strumia,
Nucl.\ Phys.\ B {\bf 458} (1996) 3;
G.~Barenboim, K.~Huitu and M.~Raidal,
Phys.\ Rev.\ D {\bf 63} (2001) 055006;
%




\bibitem{Lavignac:2001vp}
S.~Lavignac, I.~Masina and C.~A.~Savoy,
arXiv:hep-ph/0106245.
%
S.~Lavignac, I.~Masina and C.~A.~Savoy,
Nucl.\ Phys.\ B {\bf 633} (2002) 139
[arXiv:hep-ph/0202086].

\bibitem{Brooks:1999pu}
M.~L.~Brooks {\it et al.}  [MEGA Collaboration],
Phys.\ Rev.\ Lett.\  {\bf 83} (1999) 1521
[arXiv:hep-ex/9905013].


\bibitem{prop}
L.M. Barkov {\it et al.,} Research proposal for an experiment
at PSI: R-99-05.1 . \\
L. Serin, R Stroynowski, ATLAS internal note. 



\bibitem{Sato:2000zh}
J.~Sato, K.~Tobe and T.~Yanagida,
Phys.\ Lett.\ B {\bf 498} (2001) 189
[arXiv:hep-ph/0010348].



\bibitem{Santamaria:1993ah}
G.~C.~Branco, L.~Lavoura and M.~N.~Rebelo,
Phys.\ Lett.\ B {\bf 180} (1986) 264.
%
A.~Santamaria,
Phys.\ Lett.\ B {\bf 305} (1993) 90
[arXiv:hep-ph/9302301].


\bibitem{Maki:1962mu}
Z.~Maki, M.~Nakagawa and S.~Sakata,
Prog.\ Theor.\ Phys.\  {\bf 28} (1962) 870.


\bibitem{Davidson:2002qv}
M.~Fujii, K.~Hamaguchi and T.~Yanagida,
Phys.\ Rev.\ D {\bf 65} (2002) 115012
[arXiv:hep-ph/0202210]; 
%
S.~Davidson and A.~Ibarra,
Phys.\ Lett.\ B {\bf 535} (2002) 25
[arXiv:hep-ph/0202239].


\bibitem{Dighe:1999bi}
A.~S.~Dighe and A.~Y.~Smirnov,
Phys.\ Rev.\ D {\bf 62} (2000) 033007
[arXiv:hep-ph/9907423].
%
V.~Barger, D.~Marfatia and B.~P.~Wood,
Phys.\ Lett.\ B {\bf 532} (2002) 19
[arXiv:hep-ph/0202158].
%
H.~Minakata and H.~Nunokawa,
Phys.\ Lett.\ B {\bf 504} (2001) 301
[arXiv:hep-ph/0010240].


\bibitem{Olive:2000ij}
K.~A.~Olive, G.~Steigman and T.~P.~Walker,
Phys.\ Rept.\  {\bf 333} (2000) 389.


\bibitem{Plumacher:1997kc}
M.~Plumacher,
Z.\ Phys.\ C {\bf 74} (1997) 549.
M.~Plumacher,
Nucl.\ Phys.\ B {\bf 530} (1998) 207.


\bibitem{Hamaguchi:2001gw}
K.~Hamaguchi, H.~Murayama and T.~Yanagida,
Phys.\ Rev.\ D {\bf 65} (2002) 043512
[arXiv:hep-ph/0109030].



\bibitem{Covi:1996wh}
L.~Covi, E.~Roulet and F.~Vissani,
Phys.\ Lett.\ B {\bf 384} (1996) 169.



\bibitem{Kawasaki:1995af}
M.~Kawasaki and T.~Moroi,
Prog.\ Theor.\ Phys.\  {\bf 93} (1995) 879;
%
J.~R.~Ellis, G.~B.~Gelmini, J.~L.~Lopez, D.~V.~Nanopoulos and S.~Sarkar,
Nucl.\ Phys.\ B {\bf 373} (1992) 399.
%
J.~R.~Ellis, D.~V.~Nanopoulos, K.~A.~Olive and S.~J.~Rey,
Astropart.\ Phys.\  {\bf 4} (1996) 371;
%
M.~Bolz, A.~Brandenburg and W.~Buchmuller,
Nucl.\ Phys.\ B {\bf 606} (2001) 518.



\bibitem{Buchmuller:2001dc}
W.~Buchmuller and D.~Wyler,
Phys.\ Lett.\ B {\bf 521} (2001) 291
[arXiv:hep-ph/0108216].


\bibitem{Baer:2001cb}
H.~Baer, C.~Balazs, S.~Hesselbach, J.~K.~Mizukoshi and X.~Tata,
Phys.\ Rev.\ D {\bf 63} (2001) 095008
[arXiv:hep-ph/0012205].

\bibitem{Ellis:2002fe}
J.~Ellis, J.~Hisano, M.~Raidal and Y.~Shimizu,
arXiv:hep-ph/0206110.


\bibitem{RS}
{\it for recent analyses, see e.g.}
A.~Romanino and A.~Strumia,
Nucl.\ Phys.\ B {\bf 622} (2002) 73 ; [hep-ph/0108275].
S.~Abel, S.~Khalil and O.~Lebedev,
Nucl.\ Phys.\ B {\bf 606} (2001) 151 ; [hep-ph/0103320].
S.~Pokorski, J.~Rosiek and C.~A.~Savoy,
Nucl.\ Phys.\ B {\bf 570} (2000) 81 ; [hep-ph/9906206].


\bibitem{DIP} S. Davidson, A. Ibarra, M Plumacher, work in progress.



\bibitem{Barbieri:2000ma}
R.~Barbieri, P.~Creminelli, A.~Strumia and N.~Tetradis,
Nucl.\ Phys.\ B {\bf 575} (2000) 61.

\bibitem{Ellis:2002xg}
J.~Ellis and M.~Raidal,
arXiv:hep-ph/0206174.



\end{thebibliography}
\end{document}